\documentclass[a4paper,11pt]{article}
\pdfoutput=1 

\usepackage{jheppub} 

\usepackage[T1]{fontenc} 
\usepackage{orcidlink}
\usepackage{ulem}   
\usepackage{xcolor}
\usepackage{amsmath,bbm,latexsym,amssymb}
\newcommand{\Z}[1]{\ensuremath{\mathbbm{Z}_{#1}}} 
\def\vev#1{\left\langle #1\right\rangle}

\usepackage{bbold}
\usepackage{physics}
\newcommand{\one}{\mathbb{1}}
\newcommand{\AddrCFTP}{%
 Departamento de F\'\i sica and CFTP, Instituto Superior T\'ecnico\\
 Universidade de Lisboa, 
          Av. Rovisco Pais 1, 1049-001 Lisboa, Portugal }

\def\gsim{\raise0.3ex\hbox{$\;>$\kern-0.75em\raise-1.1ex\hbox{$\sim\;$}}}
\def\lsim{\raise0.3ex\hbox{$\;<$\kern-0.75em\raise-1.1ex\hbox{$\sim\;$}}}

\allowdisplaybreaks

\begin{document}


\title{Machine Learning in the 2HDM2S model for Dark Matter} 
\author[a]{Rafael Boto \orcidlink{0000-0002-9093-205X}}
\author[a]{T.~P. Rebelo \orcidlink{0009-0003-7328-3136}}
\author[a]{Jorge C.~Romão \orcidlink{0000-0002-9683-4055}}
\author[a]{João  P.~Silva \orcidlink{0000-0002-6455-9618}} 
 
\affiliation[a]{\AddrCFTP}

\emailAdd{rafael.boto@tecnico.ulisboa.pt}
\emailAdd{tiagorebelo19@tecnico.ulisboa.pt}
\emailAdd{jorge.romao@tecnico.ulisboa.pt}
\emailAdd{jpsilva@cftp.ist.utl.pt}

\abstract{
We introduce a two real scalar singlet extension of the two Higgs doublet model.
We study the vacuum structure, the bounded from below conditions, the restrictions
from the oblique parameters S,T and U, as well as the unitarity
constraints. We submit the model to collider and Dark Matter experimental
constraints and explore its allowed parameter space.
We compare randomly populated simulations,
simulations starting near the alignment limit,
and a Machine Learning based exploration.
Using Evolutionary Strategies,
we efficiently search for regions with a viable Dark Matter candidate.}

\maketitle

\section{\label{Intro}Introduction}

Despite the large success of the Standard Model (SM) of particle physics in
providing experimental predictions, leading to the discovery of a new scalar
particle resembling the predicted Higgs boson \cite{ATLAS:2012yve,CMS:2012qbp},
there is a general consensus that there must be Physics
Beyond the Standard Model (BSM).
The critical open problems consist of: the need for new sources of CP-violation,
a necessary ingredient for a successful explanation of the baryon asymmetry
of the universe~\cite{Sakharov:1967dj}; an explanation for the existence
and nature of Dark Matter (DM), which comprises of order 85 \% of the matter
content of the Universe~\cite{Planck:2013oqw}; and the origin of the observed
tiny neutrino masses.
Motivated by the lack of a fundamental reason why the scalar sector should
be limited to a single Higgs doublet, Higgs-sector extensions are required
for many of the viable explanations to these problems. 

In this work, we consider a two
Higgs doublet model (2HDM)~\cite{Branco:2011iw} with a $\Z2$ symmetry
imposed in order to forbid Higgs-mediated flavor changing
neutral couplings (FCNCs) at the tree-level, with fermions of a
given electric charge coupling to only one Higgs doublet.
Of the four possible choices, the type II 2HDM is the most studied,
since it corresponds to the structure present in supersymmetric
models - see, for example, refs.~\cite{Djouadi:2005gj,Dreiner:2023yus}.
We, however, consider the general type II 2HDM and aim to extend the scalar
sector with  additional singlets as viable candidates for particle Dark Matter.
When protected by a new global symmetry that remains unbroken in
the vacuum, the singlets can meet the requirements of heavy, stable,
electrically neutral particles of non-baryonic nature
\cite{Dimopoulos:1989hk,McDermott:2010pa}.
We will consider a scenario  where the DM particle are produced via
the freeze-out mechanism \cite{Bertone:2004pz,Feng:2010gw}.
A particle candidate with a mass similar to the scale of electroweak symmetry breaking and
an interaction cross section with the SM particles of the order of the weak
force processes can meet the observed relic abundance, and fits the class of
Weakly Interacting Massive Particles (WIMPs).
The model is then subjected to all theoretical, collider, astrophysical,
and cosmological constraints to obtain the allowed parameter space region.

The 2HDM extended by an additional real singlet with a $\Z2$ parity symmetry
may contain a viable Dark Matter candidate when the singlet does not
acquire a vacuum expectation value (vev), see
\textit{e.g.} refs.~\cite{Grzadkowski:2009iz, Logan:2010nw, Boucenna:2011hy, He:2011gc,
Bai:2012nv, Drozd:2014yla, Campbell:2015fra}.
It has been dubbed the Next-to-Two Higgs Doublet model (N2HDM) in
\cite{Chen:2013jvg, Muhlleitner:2016mzt, Ferreira:2019iqb, Engeln:2020fld, Glaus:2022rdc}.
In 2022, there was already a tension in the N2HDM between the relic density
measured by PLANCK \cite{Planck:2018vyg}
and exclusion bounds from DM direct detection experiments,
as shown in figure~3 of \cite{Glaus:2022rdc}.
Since then,
the situation has worsened,
given the great improvement in the scattering bounds \cite{LZ:2024zvo}.

A possible alternative is to consider a Type II 2HDM augmented with
a complex scalar singlet (2HDMS),
either without a DM candidate \cite{Baum:2018zhf, Heinemeyer:2021msz},
or with a DM candidate enabled through a suitable symmetry and vacuum
\cite{Dutta:2022qeq, Dutta:2023cig, Lahiri:2024rya}.
Such studies have concentrated thus far on
benchmarks rather than dedicated multi-variable scans.

Extensions of the scalar sector of the 2HDM through the addition of more
than one singlet offer promising frameworks that could also address the
Dark Matter puzzle~\footnote{The interesting possibility of adding multiple singlets to the
(one Higgs) SM, has also been considered. For a recent analysis see, for example,
Ref.~\cite{Goncalves:2025snm}.}. 
Within this context, we consider a two real scalar inert singlet extension
of the two Higgs doublet model (2HDM2S).
It includes all theoretical derivations relevant for unitarity, boundedness from below,
and vacuum stability.
It also includes a full parameter scan simulation of the model, performed in three ways:
i) with a traditional scan of the full parameter space;
ii) starting from scans close to the alignment limit;
and, iii) using a Machine Learning optimization technique,
evoking Evolutionary Strategies.

The work's structure is as follows.
In section \ref{model},
the scalar potential of the model and its particle content are developed.
In section \ref{vacuum},
the vacuum structure and the interplay of
multiple vacua is analysed.
The boundedness from below sufficient conditions,
the expressions for the potential at the minima,
the perturbative unitarity constraints,
and EW precision observables constraints for this model are developed
in sections \ref{BFB}, \ref{Global minimum}, \ref{sec:Unitarity}, and \ref{sec:STU},
respectively.
An overview of the experimental constraints applied to this model
is given in section \ref{sec:exp}.
In section \ref{sec:samp},
the computational methods used are explained,
and a comparison of results between different approaches is done.
A discussion of results satisfying all described constraints is
provided in section \ref{sec:results},
followed by conclusions in section \ref{sec:conclusion}.
The appendix shows a comparison between the desired vacuum and all other
possible vacua within this model.

\section{\label{model}The Scalar Potential of the 2HDM2S}

Our goal is to analyze an extension to the SM, with two doublets and two real scalars, in which the scalars do not acquire a Vacuum Expectation Value (vev) after Spontaneous Symmetry Breaking (SSB). 
In order to avoid dangerous flavor-changing neutral couplings (FCNC) at the tree level, a $\Z2$ symmetry softly broken by the $m_{12}^2$ term, to introduce a decoupling limit~\cite{Gunion:2002zf}, is imposed on the Lagrangian. In
addition, we assume CP conservation in the theory; therefore, all the coefficients in the
potential are taken to be real. 

The 2HDM2S contains an additional complex doublet to the Standard Model, $\Phi_2$, and two real scalars, S and P, giving rise to additional
terms in the scalar potential, including one $\mathbb{Z}'_{2}$
symmetry imposed on the real
scalar singlet fields S and P.
The imposed symmetries are
\begin{equation}
    \mathbb{Z}_2 : \;\;\;\; \Phi_1 \rightarrow \Phi_1, \;\;\;\; \Phi_2 \rightarrow - \Phi_2, \;\;\;\; S \rightarrow S, \;\;\;\; P \rightarrow P ,
\end{equation}
and
\begin{equation}
    \mathbb{Z}_2' : \;\;\;\; \Phi_1 \rightarrow \Phi_1, \;\;\;\; \Phi_2 \rightarrow  \Phi_2, \;\;\;\; S \rightarrow - S, \;\;\;\; P \rightarrow - P .
\end{equation}

The terms invariant under the symmetry are written as,
\begin{equation}
\begin{aligned}
    V = & \, m_{11}^2 (\Phi_1^\dagger \Phi_1) + m_{22}^2 (\Phi_2^\dagger \Phi_2) 
    - \big[ m_{12}^2 (\Phi_1^\dagger \Phi_2) + \text{h.c.} \big] 
     + \frac{\lambda_1}{2} (\Phi_1^\dagger \Phi_1)^2 
    + \frac{\lambda_2}{2} (\Phi_2^\dagger \Phi_2)^2 \\ 
    & + \lambda_3 (\Phi_1^\dagger \Phi_1)(\Phi_2^\dagger \Phi_2) 
    + \lambda_4 (\Phi_1^\dagger \Phi_2)(\Phi_2^\dagger \Phi_1) 
    + \bigg[ \frac{\lambda_5}{2} (\Phi_1^\dagger \Phi_2)^2 + \text{h.c.} \bigg] \\
    & \, \frac{1}{2} m_S^2 S^2 + \frac{1}{2} m_P^2 P^2 - m_{SP}^2 S P +\frac{1}{8} \lambda_6 S^4 + \frac{1}{8} \lambda_9 P^4 + \frac{1}{4} \lambda_{10} S^2 P^2 
     +\frac{1}{6} \lambda_{13} S^3 P  \\
     & + \frac{1}{6} \lambda_{14} S P^3 + \frac{1}{2} \big( \lambda_7 (\Phi_1^\dagger \Phi_1) + \lambda_8 (\Phi_2^\dagger \Phi_2) \big) S^2 
    + \frac{1}{2} \big( \lambda_{11} (\Phi_1^\dagger \Phi_1)  +\lambda_{12} (\Phi_2^\dagger \Phi_2) \big) P^2 \\
    & + \frac{1}{2} \big( \lambda_{15} (\Phi_1^\dagger \Phi_1) + \lambda_{16} (\Phi_2^\dagger \Phi_2) \big) S P.
    \label{2hdm2s}
\end{aligned}
\end{equation}

We can alternatively write potential of the 2HDM2S as,
\begin{equation}
  \label{eq:1}
  V= V_2 + V_4\ ,
\end{equation}
where the quadratic part is
\begin{equation}
  \label{eq:2}
  V_2=m_{11}^2 (\Phi_1^\dagger \Phi_1) + m_{22}^2  (\Phi_2^\dagger
  \Phi_2) - \left( m_{12}^2  (\Phi_1^\dagger \Phi_2) +
    \text{h.c.}\right)
  + \frac{1}{2} m_S^2 S^2 + \frac{1}{2} m_P^2 P^2 - m_{SP}^2 S P \, .
\end{equation}
and the quartic terms are,
\begin{equation}
  V_{\rm Quartic}=V_N + V_{CB} + V_{HC} + V_{O}\, ,
\end{equation}
where
\begin{align}
V_N=& \frac{\lambda_1}{2} (\Phi_1^\dagger \Phi_1)^2 +
\frac{\lambda_2}{2} (\Phi_2^\dagger \Phi_2)^2 +
(\lambda_3+\lambda_4) (\Phi_1^\dagger \Phi_1)(\Phi_2^\dagger \Phi_2) + \frac{1}{8} \lambda_6 S^4 + \frac{1}{8}\lambda_9 P^4
\nonumber\\[+2mm]
&
+ \frac{1}{4}\lambda_{10} S^2 P^2
+ \frac{1}{2}\left( \lambda_7(\Phi_1^\dagger \Phi_1) +
\lambda_8 (\Phi_2^\dagger \Phi_2)\right) S^2 +
\frac{1}{2}\left( \lambda_{11}(\Phi_1^\dagger \Phi_1)+
\lambda_{12} (\Phi_2^\dagger \Phi_2)\right) P^2 \, ,\\[+2mm]
  V_{CB}=&-\lambda_4 z_{12} \, ,\\[+2mm]
V_{HC}=& \frac{1}{2}\lambda_5 \left(  (\Phi_1^\dagger \Phi_2)^2 + \text{h.c.} \right) \, ,\\[+2mm]
V_{O}=&
\frac{1}{6}\lambda_{13} S^3 P + 
\frac{1}{6}\lambda_{14}S P^3 +
\frac{1}{2}\left( \lambda_{15}(\Phi_1^\dagger \Phi_1)+
\lambda_{16} (\Phi_2^\dagger \Phi_2)\right) S P \, ,
\end{align}
where we have defined~\cite{Faro:2019vcd},
\begin{eqnarray}
  z_{12}= (\Phi_1^\dagger \Phi_1)(\Phi_2^\dagger \Phi_2)-
  (\Phi_1^\dagger \Phi_2)(\Phi_2^\dagger \Phi_1) .
\end{eqnarray}

We consider the vacuum where the singlets do \textit{not} develop a vev,
thus keeping the $\mathbb{Z}_2'$ symmetry intact.
After SSB, we can develop the doublets and singlets in the symmetry basis as:
\begin{equation}
    \Phi_1 = 
\begin{pmatrix}
w_1^+ \\
\frac{(v_1+y_1 + i z_1)}{\sqrt{2}} 
\end{pmatrix}, 
\;\;\;\;
 \Phi_2  = 
\begin{pmatrix}
w_2^+ \\
\frac{(v_2+y_2 + i z_2)}{\sqrt{2}}
\end{pmatrix}, 
\;\;\;\;
 S  = s,
 \;\;\;\;
  P = p .
\end{equation}
The particle content of the model consists of four CP-even Higgs bosons, $h_i$, $i\in \{1,2,3,4\}$, where $h_3$ and $h_4$ are on the dark sector, a pseudoscalar Higgs, $A$, and a charged Higgs $H^\pm$. 

Due to charge and CP conservation and the imposed symmetries, the $8 \times 8$ mass matrix is decomposed into four blocks: the $2 \times 2$ matrix for the charged fields, the $2 \times 2$ matrix for the CP-odd fields and two $2 \times 2$ matrices for the CP-even states, since the fields from the doublets do not mix with the scalars $s$ and $p$. The charged and pseudoscalar sectors remain unchanged in regard to the 2HDM. Hence, the charged and pseudoscalar mass matrices may be diagonalized by the rotation matrix
\begin{equation}\label{obeta}
    \mathcal{O}_{\beta} =
    \begin{pmatrix}
        c_{\beta} &  s_{\beta} \\ - s_{\beta} & c_{\beta}
    \end{pmatrix},
\end{equation}
where $\tan\beta= \frac{v_2}{v_1}$, $s_\beta=\sin(\beta)$, and
$c_\beta=\cos(\beta)$.
With this definition we may obtain the physical-mass eigenstates~\cite{Das:2019yad} by
\begin{equation}
\begin{bmatrix}G^0\\A\end{bmatrix}=\mathcal{O}_{\beta}\begin{bmatrix}z_1\\z_2\end{bmatrix},\quad
\begin{bmatrix}G^+\\H^+\end{bmatrix}=\mathcal{O}_{\beta}\begin{bmatrix}w_1^+\\w_2^+\end{bmatrix},
\end{equation}
where $G^+$ and $G^0$ are the Nambu-Goldstone bosons.
For the scalar states, we obtain

\begin{equation}\label{field_eq}
\begin{bmatrix}h_1\\h_2\end{bmatrix} = 
\begin{bmatrix}
c_{\alpha_1} & -s_{\alpha_1} \\
s_{\alpha_1} & c_{\alpha_1}
\end{bmatrix}
\begin{bmatrix}y_1\\y_2\end{bmatrix}, \\
\begin{bmatrix}h_3\\h_4\end{bmatrix} = 
\begin{bmatrix}
c_{\alpha_2} & -s_{\alpha_2} \\
s_{\alpha_2} & c_{\alpha_2}
\end{bmatrix}
\begin{bmatrix}s\\p\end{bmatrix}.
\end{equation}
Fixing the mass basis, one derives the corresponding 22 free parameters:
\begin{equation}
    m_{h1, h2, h3, h4},\; m_A,\; m_{H\pm},\; m_{12}^2,\; m_S^2, \; m_P^2, \; m_{SP}^2,\; \lambda_{6, 8, 9, 10, 12, 13, 14, 16},\;\alpha_{1, 2},\; t_{\beta},\; v\;  . 
\end{equation}
We may now write the quartic couplings $\lambda_i$ in terms of the physical basis, obtaining the following expressions:
\begin{align}
    \tilde{\mu}^2 =& \frac{m_{12}^2}{s_\beta c_\beta}, 
    \\
    \lambda_1 =& \frac{1}{v^2 c_{\beta}^2} \big( m_1^2 c_{\alpha_1}^2 + m_2^2 s_{\alpha_1}^2 - \tilde{\mu}^2 s_{\beta}^2  \big) ,
    \\
    \lambda_2 =& \frac{1}{v^2 s_{\beta}^2} \big(m_1^2 s_{\alpha_1}^2 + m_2^2 c_{\alpha_1}^2 - \tilde{\mu}^2 c_{\beta}^2  \big) ,
    \\
    \lambda_3 =& \frac{1}{v^2} \big(\frac{s_{\alpha_1} c_{\alpha_2}} {s_{\beta} c_{\beta}}\big( m_2^2 - m_1^2\big) - \tilde{\mu}^2 + 2 m^2_{H\pm}  \big),
    \\
    \lambda_4 =& \frac{1}{v^2} \big(\tilde{\mu}^2 + m_A^2 - 2 m^2_{H\pm} \big),
    \\
    \lambda_5 =& \frac{1}{v^2} \big(\tilde{\mu}^2 - m_A^2 \big),
    \\
    \lambda_7 =& \frac{1}{v^2 c_{\beta}^2} \big( 2 m_{DMS}^2 c_{\alpha_2}^2 + 
 2 m_{DMP}^2 s_{\alpha_2}^2 - 2 m_S^2 - \lambda_8 v^2 s_{\beta}^2\big) ,
    \\
     \lambda_{11} =& \frac{1}{v^2 c_{\beta}^2} \big( 2 m_{DMS}^2 s_{\alpha_2}^2 + 2 m_{DMP}^2 c_{\alpha_2}^2 - 2 m_P^2 - \lambda_{12} v^2 s_{\beta}^2\big) ,
     \\
     \lambda_{15} =& \frac{1}{v^2 c_{\beta}^2} \big[ 4 \big( m_{DMP}^2 - m_{DMS}^2 \big) s_{\alpha_2} c_{\alpha_2} + 4 m_{SP}^2 - \lambda_{16} v^2 s_{\beta}^2\big] .
\end{align}

\section{\label{vacuum} Vacuum Structure}

\subsection{\label{stationarity} Stationarity Conditions}

The most general constant field configuration for the vacuum is, using the notation from~\cite{Muhlleitner:2016mzt, Ferreira:2019iqb},

\begin{equation}
    \langle \Phi_1 \rangle = \frac{1}{\sqrt{2}}
\begin{pmatrix}
0 \\
v_1
\end{pmatrix}, 
\;\;\;\;
 \langle \Phi_2 \rangle = \frac{1}{\sqrt{2}}
\begin{pmatrix}
v_{cb} \\
v_2 + i v_{cp}
\end{pmatrix}, 
\;\;\;\;
 \langle S \rangle = v_S,
 \;\;\;\;
 \langle P \rangle = v_P .
\end{equation}

In order to find all the possible minima, we consider the following stationarity conditions for the vevs or, equivalently, for the complex charged fields $w_i^+ \; (i \in \{1,2\})$, the real neutral CP-even, $y_1, y_2, s$, and $p$, and CP-odd fields, $z_1$, and $z_2$ : 
\begin{align}
     \left\langle \frac{\partial V}{\partial w_1^+} \right\rangle = 0 
     \iff & v_{cb} \big(v_1 v_2 \big( \lambda_4 + \lambda_5 \big) - 2 m_{12}^2 \big) = 0, \label{eq: charged1}
    \\
       \left\langle \frac{\partial V}{\partial w_2^+} \right\rangle = \left\langle \frac{\partial V}{\partial v_{cb}} \right\rangle = 0 
    \iff & -v_{cb} m_{12}^2 
    = \frac{1}{2} v_{cb} \big(v_1^2 \lambda_3 + v_2^2 \lambda_3 + v_{cb}^2 \lambda_2 + v_{cp}^2 \lambda_2 + v_S^2 \lambda_8 \notag \\
   &+ v_P^2 \lambda_{12} + v_P v_S \lambda_{16} \big),
    \label{eq:charged2/vcb} 
    \\
    \left\langle \frac{\partial V}{\partial y_1} \right\rangle = \left\langle \frac{\partial V}{\partial v_1} \right\rangle =0 
    \iff &v_2 m_{12}^2 - v_1 m_1^2 = \frac{1}{2} v_1 \big(v_1^2 \lambda_1 + v_2^2 \lambda_{345} + v_{cb}^2 \lambda_3 + v_{cp}^2 \lambda_{34-5} \notag \\
    &+ v_S^2 \lambda_7 + v_P^2 \lambda_{11} + v_S v_P \lambda_{15} \big), \label{eq:v1} 
    \\
    \left\langle \frac{\partial V}{\partial y_2} \right\rangle = \left\langle \frac{\partial V}{\partial v_2} \right\rangle =0 
    \iff &v_1 m_{12}^2 - v_2 m_2^2 = \frac{1}{2} v_2 \big(v_1^2 \lambda_{345} + v_2^2 \lambda_2 + v_{cb}^2 \lambda_2 + v_{cp}^2 \lambda_2 \notag \\
    &+ v_S^2 \lambda_8 + v_P^2 \lambda_{12} + v_S v_P \lambda_{16} \big), \label{eq:v2} 
    \\
    \left\langle \frac{\partial V}{\partial z_1} \right\rangle = 0 
     \iff & v_{cp} \big(v_1 v_2 \lambda_5 - 2 m_{12}^2 \big) = 0, \label{eq:eta1}
     \\
    \left\langle \frac{\partial V}{\partial z_2} \right\rangle = \left\langle \frac{\partial V}{\partial v_{cp}} \right\rangle = 0 
    \iff &-v_{cp} m_{12}^2 = \frac{1}{2} v_{cp} \big(v_1^2 \lambda_{34-5} + v_2^2 \lambda_2 + v_{cb}^2 \lambda_2 + v_{cp}^2 \lambda_2 \notag \\
    &+ v_S^2 \lambda_8 + v_P^2 \lambda_{12} + v_S v_P \lambda_{16} \big), \label{eq:vcp/eta2}
    \\
    \left\langle \frac{\partial V}{\partial s} \right\rangle = \left\langle \frac{\partial V}{\partial v_S} \right\rangle =0 
    \iff & v_P m_{SP}^2 - v_S m_S^2 = \frac{1}{2} v_S \big(v_1^2 \lambda_7 + v_2^2 \lambda_8 + v_{cb}^2 \lambda_8 + v_{cp}^2 \lambda_8 \notag \\
    & + v_S^2 \lambda_6 + v_P^2 \lambda_{10} + v_S v_P \lambda_{13} \big) + \frac{1}{4} v_P \big(v_1^2 \lambda_{15} + v_2^2 \lambda_{16} \notag \\
    & + v_{cb}^2 \lambda_{16} + v_{cp}^2 \lambda_{16}\big) + \frac{1}{6} v_P^3 \lambda_{14}, \label{rhoS/vS}
\end{align}
\begin{align}
    \left\langle \frac{\partial V}{\partial p} \right\rangle = \left\langle \frac{\partial V}{\partial v_P} \right\rangle =0 
    \iff & v_S m_{SP}^2 - v_P m_P^2 = \frac{1}{2} v_P \big(v_1^2 \lambda_{11} + v_2^2 \lambda_{12} + v_{cb}^2 \lambda_{12} + v_{cp}^2 \lambda_{12} \notag \\
    & + v_S^2 \lambda_{10} + v_P^2 \lambda_9 + v_S v_P \lambda_{14} \big) + \frac{1}{4} v_S \big(v_1^2 \lambda_{15} + v_2^2 \lambda_{16} \notag \\
    & + v_{cb}^2 \lambda_{16} + v_{cp}^2 \lambda_{16}\big) + \frac{1}{6} v_S^3 \lambda_{13}, \label{rhoP/vP}
\end{align}
where
\begin{equation}
    \lambda_{345} \equiv \lambda_3 + \lambda_4 + \lambda_5 ,\;\;\;\;
    \lambda_{34-5} \equiv \lambda_3 + \lambda_4 - \lambda_5 .
\end{equation}
From eqs.~(\ref{eq:charged2/vcb}) and (\ref{eq:vcp/eta2}) we conclude that
\begin{equation}
    \big(\lambda_4 = \lambda_5) \vee \big[\big(v_{cb} \neq 0 \Rightarrow v_{cp} = 0 \big) \vee \big(v_{cp} \neq 0 \Rightarrow v_{cb} = 0 \big) \big].
\end{equation}
From eqs.~(\ref{eq:v1}) and (\ref{eq:v2}) we observe that
\begin{equation}
    \big( v_1 = 0 \Leftrightarrow v_2 = 0 \big) \vee m_{12}^2 = 0.
\end{equation}
From eqs.~(\ref{eq: charged1}) and (\ref{eq:eta1}), we further infer that
\begin{equation}
    v_1 = v_2 = 0 \Rightarrow \big( v_{cb} = v_{cp} = 0 \vee m_{12}^2 = 0 \big).
\end{equation}
These results are similar to those on N2HDM.

From the eigenvalues of the scalar mass matrices, we derive, at a given stationary point $i$, the squared charged scalar mass and  the pseudoscalar squared mass (for $v_{cb} = 0 \wedge v_{vp} = 0$)
\begin{equation}
    \big(m_{H\pm}^2 \big)_i = m_{12}^2 \frac{v_i^2}{(v_1)_i (v_2)_i} - \frac{1}{2} (\lambda_4 + \lambda_5) v_i^2,  \;\;\; \big(m_A^2 \big)_i = \big(m_{H\pm}^2 \big)_i + \frac{1}{2} (\lambda_4 - \lambda_5) v_i^2.
\end{equation}
From the scalar mass matrix of the singlets, we define the squared mass of the singlet field S and P without mass mixing as $\big( m_s^2 \big)_i$ and $\big( m_p^2 \big)_i$ which, due to the required copositivity of the CP-even mass matrix, are positive. For the mass mixing term (off-diagonal), we define it as $\big(m_{sp}^2 \big)_i$, which is not necessarily positive.

\subsection{\label{Formalism and Vacua}Formalism and Vacua}

To study the interplay of multiple vacua, it is useful to
introduce a bilinear formalism, similar to the one used
in~\cite{Lahiri:2024rya, Ferreira:2019iqb} for the 2HDMS and N2HDM.
The relevant bilinears for this work are the following.
\begin{equation}
    x_1 = |\Phi_1|^2,  x_2 = |\Phi_2|^2,  x_3 = \operatorname{Re}\big(\Phi_1^{\dagger} \Phi_2\big),  x_4 = \operatorname{Im}\big(\Phi_1^{\dagger} \Phi_2\big),  x_5 = \frac{1}{2} S^2,  x_6 = \frac{1}{2} P^2,  x_7 = \frac{1}{2} S P .
\end{equation}
We define the vectors X, A and the symmetric matrix B as
\begin{equation}
    X = 
    \begin{pmatrix}
x_1 \\ x_2 \\ x_3 \\ x_4 \\ x_5 \\ x_6 \\ x_7
\end{pmatrix} ,
\;\;
A =
    \begin{pmatrix}
m_{11}^2 \\ m_{22}^2 \\ -2 m_{12}^2 \\ 0 \\ m_S^2 \\ m_P^2 \\ -2 m_{SP}^2
\end{pmatrix},
\;\;
B =
\begin{pmatrix}
    \lambda_1 & \lambda_3 & 0 & 0 & \lambda_7 & \lambda_{11} & \lambda_{15} \\
    \lambda_3 & \lambda_2 & 0 & 0 & \lambda_8 & \lambda_{12} & \lambda_{16} \\
    0 & 0 & 2(\lambda_4 + \lambda_5) & 0 & 0 & 0 & 0 \\
    0 & 0 & 0 & 2(\lambda_4 - \lambda_5) & 0 & 0 & 0 \\
    \lambda_7 & \lambda_8 & 0 & 0 & \lambda_6 & \lambda_{10} & \frac{2}{3}\lambda_{13} \\
    \lambda_{11} & \lambda_{12} & 0 & 0 & \lambda_{10} & \lambda_9 & \frac{2}{3}\lambda_{14} \\
    \lambda_{15} & \lambda_{16} & 0 & 0 & \frac{2}{3}\lambda_{13} & \frac{2}{3}\lambda_{14} & 0 
\end{pmatrix} .
\end{equation}
In terms of the bilinears, the potential may be written as
\begin{equation}
    V = A^T X + \frac{1}{2} X^T B X .
\end{equation}
We also make use of the vector
\begin{equation}
    V' = \frac{\partial V}{\partial X^T} = A + B X,
\end{equation}
so that, at a stationary point $i$, the value of the potential is given by 
\begin{equation}
    V_i = \frac{1}{2} A^T \langle X \rangle _i = - \frac{1}{2} \langle X \rangle _i ^T B \langle X \rangle _i. \label{stat_eq}
\end{equation}

The procedure for finding the stability conditions between two stationary points $i$ and $j$ is as follows. The internal product between $X_i$ and $V'_j$ yields
\begin{equation}
    \langle X \rangle _i^T V'_j = \langle X \rangle _i^T A + \langle X \rangle_i^T B \langle X \rangle_j \, , \label{stateq_i}
\end{equation}
and the internal product between $X_j$ and $V'_i$ results in
\begin{equation}
    \langle X \rangle _j^T V'_i = \langle X \rangle _j^T A + \langle X \rangle_j^T B \langle X \rangle_i . \label{stateq_j}
\end{equation}
From eq.~(\ref{stat_eq}), one can write
\begin{equation}
    \langle X \rangle _i^T A = 2 V_i, \;\;\;\; \langle X \rangle _j^T A = 2 V_j \, ,
\end{equation}
and, since $B$ is symmetric, combining eqs.~\eqref{stateq_i} and \eqref{stateq_j} results in
\begin{equation}\label{min-comparison}
    V_i - V_j = \frac{1}{2} \big(\langle X \rangle _i^T V'_j - \langle X \rangle_j^T V'_i \big) .
\end{equation}

Due to the results from the minimization conditions, this model has four possible EW vacua, $\mathcal{N}, \mathcal{N}_s, \mathcal{N}_p$ and $\mathcal{N}_{sp}$, four possible charge-breaking vacua, $\mathcal{CB},\mathcal{CB}_s, \mathcal{CB}_p$ and $\mathcal{CB}_{sp}$, four possible $\mathcal{CP}$-breaking vacua, $\mathcal{CP}, \mathcal{CP}_s, \mathcal{CP}_p$ and $\mathcal{CP}_{sp}$, and 3 neutral vacua, $\mathcal{S}, \mathcal{P}$, and $\mathcal{SP}$. We define them as follows.
\begin{align}
\mathcal{N} \to \langle \Phi_1 \rangle_0 &= \frac{1}{\sqrt{2}} 
\begin{pmatrix} 
0 \\ 
v_1 
\end{pmatrix},
\quad \langle \Phi_2 \rangle_0 = \frac{1}{\sqrt{2}} 
\begin{pmatrix} 
0 \\ 
v_2 
\end{pmatrix}, 
\quad \langle S \rangle_0 = 0, 
\quad \langle P \rangle_0 = 0,
\label{vev_nosso}
\\ 
\mathcal{N}_s \to \langle \Phi_1 \rangle_0 &= \frac{1}{\sqrt{2}} 
\begin{pmatrix} 
0 \\ 
v'_1 
\end{pmatrix},
\quad \langle \Phi_2 \rangle_0 = \frac{1}{\sqrt{2}} 
\begin{pmatrix} 
0 \\ 
v'_2 
\end{pmatrix}, 
\quad \langle S \rangle_0 = v'_S, 
\quad \langle P \rangle_0 = 0, \\ 
\mathcal{N}_p \to \langle \Phi_1 \rangle_0 &= \frac{1}{\sqrt{2}} 
\begin{pmatrix} 
0 \\ 
v''_1 
\end{pmatrix},
\quad \langle \Phi_2 \rangle_0 = \frac{1}{\sqrt{2}} 
\begin{pmatrix} 
0 \\ 
v''_2
\end{pmatrix}, 
\quad \langle S \rangle_0 = 0, 
\quad \langle P \rangle_0 = v''_P, \\ 
\mathcal{N}_{sp} \to \langle \Phi_1 \rangle_0 &= \frac{1}{\sqrt{2}} 
\begin{pmatrix} 
0 \\ 
v'''_1 
\end{pmatrix},
\quad \langle \Phi_2 \rangle_0 = \frac{1}{\sqrt{2}} 
\begin{pmatrix} 
0 \\ 
v'''_2 
\end{pmatrix}, 
\quad \langle S \rangle_0 = v'''_S, 
\quad \langle P \rangle_0 = v'''_P, \\
\mathcal{CB} \to \langle \Phi_1 \rangle_0 &= \frac{1}{\sqrt{2}} 
\begin{pmatrix} 
0 \\ 
c_1 
\end{pmatrix}, 
\quad \langle \Phi_2 \rangle_0 = \frac{1}{\sqrt{2}} 
\begin{pmatrix} 
c_2 \\ 
c_3 
\end{pmatrix}, 
\quad \langle S \rangle_0 = 0, 
\quad \langle P \rangle_0 = 0, \\ 
\mathcal{CB}_s \to \langle \Phi_1 \rangle_0 &= \frac{1}{\sqrt{2}} 
\begin{pmatrix} 
0 \\ 
c'_1 
\end{pmatrix}, 
\quad \langle \Phi_2 \rangle_0 = \frac{1}{\sqrt{2}} 
\begin{pmatrix} 
c'_2 \\ 
c'_3 
\end{pmatrix}, 
\quad \langle S \rangle_0 = c'_S, 
\quad \langle P \rangle_0 = 0, \\
\mathcal{CB}_p \to \langle \Phi_1 \rangle_0 &= \frac{1}{\sqrt{2}} 
\begin{pmatrix} 
0 \\ 
c''_1 
\end{pmatrix}, 
\quad \langle \Phi_2 \rangle_0 = \frac{1}{\sqrt{2}} 
\begin{pmatrix} 
c''_2 \\ 
c''_3 
\end{pmatrix}, 
\quad \langle S \rangle_0 = 0, 
\quad \langle P \rangle_0 = c''_P, \\ 
\mathcal{CB}_{sp} \to \langle \Phi_1 \rangle_0 &= \frac{1}{\sqrt{2}} 
\begin{pmatrix} 
0 \\ 
c'''_1 
\end{pmatrix}, 
\quad \langle \Phi_2 \rangle_0 = \frac{1}{\sqrt{2}} 
\begin{pmatrix} 
c'''_2 \\ 
c'''_3 
\end{pmatrix}, 
\quad \langle S \rangle_0 = c'''_S, 
\quad \langle P \rangle_0 = c'''_P, \\
\mathcal{CP} \to \langle \Phi_1 \rangle_0 &= \frac{1}{\sqrt{2}} 
\begin{pmatrix} 
0 \\ 
\bar{v}_1 
\end{pmatrix}, 
\quad \langle \Phi_2 \rangle_0 = \frac{1}{\sqrt{2}} 
\begin{pmatrix} 
0 \\ 
\bar{v}_2 + i \bar{v}_3  
\end{pmatrix}, 
\quad \langle S \rangle_0 = 0, 
\quad \langle P \rangle_0 = 0, \\ 
\mathcal{CP}_s \to \langle \Phi_1 \rangle_0 &= \frac{1}{\sqrt{2}} 
\begin{pmatrix} 
0 \\ 
\bar{v}'_1 
\end{pmatrix}, 
\quad \langle \Phi_2 \rangle_0 = \frac{1}{\sqrt{2}} 
\begin{pmatrix} 
0 \\ 
\bar{v}'_2 + i \bar{v}'_3 
\end{pmatrix}, 
\quad \langle S \rangle_0 = \bar{v}'_S, 
\quad \langle P \rangle_0 = 0, 
\end{align}
\begin{align}
\mathcal{CP}_p \to \langle \Phi_1 \rangle_0 &= \frac{1}{\sqrt{2}} 
\begin{pmatrix} 
0 \\ 
\bar{v}''_1 
\end{pmatrix}, 
\quad \langle \Phi_2 \rangle_0 = \frac{1}{\sqrt{2}} 
\begin{pmatrix} 
0 \\ 
\bar{v}''_2 + i \bar{v}''_3 
\end{pmatrix}, 
\quad \langle S \rangle_0 = 0, 
\quad \langle P \rangle_0 = \bar{v}''_P, \\ 
\mathcal{CP}_{sp} \to \langle \Phi_1 \rangle_0 &= \frac{1}{\sqrt{2}} 
\begin{pmatrix} 
0 \\ 
\bar{v}'''_1 
\end{pmatrix}, 
\quad \langle \Phi_2 \rangle_0 = \frac{1}{\sqrt{2}} 
\begin{pmatrix} 
0 \\ 
\bar{v}'''_2 + i \bar{v}'''_3 
\end{pmatrix}, 
\quad \langle S \rangle_0 = \bar{v}'''_S, 
\quad \langle P \rangle_0 = \bar{v}'''_P,
\\
\mathcal{S} \to \langle \Phi_1 \rangle_0 &= \frac{1}{\sqrt{2}} 
\begin{pmatrix} 
0 \\ 
0 
\end{pmatrix}, 
\quad \langle \Phi_2 \rangle_0 = \frac{1}{\sqrt{2}} 
\begin{pmatrix} 
0 \\ 
0 
\end{pmatrix}, 
\quad \langle S \rangle_0 = S, 
\quad \langle P \rangle_0 = 0, \\ 
\mathcal{P} \to \langle \Phi_1 \rangle_0 &= \frac{1}{\sqrt{2}} 
\begin{pmatrix} 
0 \\ 
0 
\end{pmatrix}, 
\quad \langle \Phi_2 \rangle_0 = \frac{1}{\sqrt{2}} 
\begin{pmatrix} 
0 \\ 
0 
\end{pmatrix}, 
\quad \langle S \rangle_0 = 0, 
\quad \langle P \rangle_0 = P, \\ 
\mathcal{SP} \to \langle \Phi_1 \rangle_0 &= \frac{1}{\sqrt{2}} 
\begin{pmatrix} 
0 \\ 
0 
\end{pmatrix}, 
\quad \langle \Phi_2 \rangle_0 = \frac{1}{\sqrt{2}} 
\begin{pmatrix} 
0 \\ 
0 
\end{pmatrix}, 
\quad \langle S \rangle_0 = S, 
\quad \langle P \rangle_0 = P.
\end{align}
The vacuum in which we focus our analysis is the $\mathcal{N}$-type vacuum
in eq.~\eqref{vev_nosso}.

\section{Sufficient BFB conditions}
\label{BFB}

We now will use a parameterization inspired by Ref.~\cite{Faro:2019vcd},
\begin{equation}
  \label{eq:22}
  \Phi_1=\sqrt{r_1}
  \begin{pmatrix}
    0\\
    1
  \end{pmatrix},\quad
  \Phi_2=\sqrt{r_2}
  \begin{pmatrix}
    \sin(\alpha_2)\\
    \cos(\alpha_2) e^{i \beta_2}
  \end{pmatrix},\quad
  S=\eta_S \sqrt{r_3},\quad
  P=\eta_P\sqrt{r_4} \, ,
\end{equation}
where $\eta_{S,P}= \pm 1$ and $r_i\geq 0$.
With this parameterization we can show that $V_N$ can be written as a quadratic form,
\begin{equation}
  V_N= \frac{1}{2} \sum_{ij} r_i A_{ij} r_j\, ,
\end{equation}
with the matrix $A$ given by
\begin{equation}
  A=\begin{bmatrix}
    \lambda_1 &\lambda_3+\lambda_4 & \frac{1}{2}
    \lambda_7&\frac{1}{2}\lambda_{11}\\
    \lambda_3+\lambda_4& \lambda_2 &\frac{1}{2}
    \lambda_8 &\frac{1}{2}\lambda_{12}\\
    \frac{1}{2}\lambda_7& \frac{1}{2}\lambda_8 & \frac{1}{4}\lambda_6 &\frac{1}{4}\lambda_{10}\\
    \frac{1}{2}\lambda_{11}&\frac{1}{2}
    \lambda_{12}&\frac{1}{4}\lambda_{10} &
    \frac{1}{4}\lambda_9
  \end{bmatrix} \, .
\end{equation}
This part of the potential is BFB if this form is positive definite
for $r_i\geq 0$.

The problem is that we have to find the conditions for a matrix of
order four to be definite positive. For the case of matrices of order
three there is a simple result. A $3\times 3$ symmetric matrix is
positive definite if and only if the following conditions, known as
copositivity conditions~\cite{Klimenko:1984qx,Kannike:2012pe}, are
satisfied:
\begin{align}
  \label{eq:5}
  & A_{11} \ge 0, A_{22} \ge 0, A_{33} \ge 0\, ,\nonumber\\[+2mm]
  &\overline{A}_{12}=\sqrt{A_{11}A_{22}} + A_{12} \ge 0,\quad
  \overline{A}_{13}=\sqrt{A_{11}A_{33}} + A_{13} \ge 0,\quad
  \overline{A}_{23}=\sqrt{A_{22}A_{33}} + A_{23} \ge
  0\, ,\nonumber\\[+2mm]
  &\sqrt{A_{11}A_{22}A_{33}} + A_{12}\sqrt{A_{33}}
  + A_{13}\sqrt{A_{22}}+ A_{23}\sqrt{A_{11}}
  +\sqrt{2 \overline{A}_{12}\overline{A}_{13}\overline{A}_{23}} \ge 0 .
\end{align}

We will come back to these expressions after we find a quadratic form that
bounds the potential from below.
As the odd part, $V_O$, cannot be written as a quadratic form,
one has to find sufficient although not necessary conditions for
the BFB~\cite{Boto:2022uwv}.
For this, we bound each part of the potential. We have
\begin{equation}
  V_{CB} \geq V_{CB}^{\rm lower}= r_1 r_2\, \text{min}(0,-\lambda_4) \, ,
\end{equation}
where we have used 
\begin{eqnarray}
  0\leq z_{12}\leq 1 \, .
\end{eqnarray}
For $V_{HC}$ we get
\begin{equation}
  V_{HC} \geq V_{HC}^{\rm lower}=-|\lambda_5| r_1 r_2  \, .
\end{equation}
Finally the odd part, $V_O$ is the more complicated. We have
\begin{equation}
 V_O= \frac{1}{6}\lambda_{13} r_3 \eta_S \eta_P \sqrt{r_3} \sqrt{r_4}
+  \frac{1}{6}\lambda_{14} r_4 \eta_S \eta_P \sqrt{r_3} \sqrt{r_4}
+\frac{1}{2} \left( \lambda_{15} r_1+\lambda_{16} r_2
\right) \eta_S \eta_P \sqrt{r_3}\sqrt{r_4}  \, .
\end{equation}

Let us consider the first term.  As all the $r_i$ are positive
definite, the worst situation occurs when $ \lambda_{13} \eta_S \eta_P
=-|\lambda_{13}|$. Then certainly we have,
\begin{equation}
  V_O^{\rm 1st} \geq -\frac{1}{6}\,|\lambda_{13}| r_3 \sqrt{r_3}\sqrt{r_4}  \; .
\end{equation}
Now we use the relation (for $r_i\geq 0$)
\begin{equation}
  - \sqrt{r_3} \sqrt{r_4} \geq -r_3 - r_4 \;,
\end{equation}
to obtain
\begin{equation}
  V_O^{\rm 1st} \geq - \frac{1}{6}\, |\lambda_{13}| r_3 (r_3 + r_4) \;,
\end{equation}
so we can safely say that
\begin{equation}
  V_O^{\rm 1st} \geq - \frac{1}{6}\, |\lambda_{13}| (r_3^3 + r_3 r_4) \;,
\end{equation}
which is a quadratic form. Continuing with this reasoning, we obtain
\begin{align}
  V_O \geq &V_O^{\rm lower}= 
  -\frac{1}{6}\, |\lambda_{13}| (r_3^2 + r_3 r_4)-
  \frac{1}{6}\, |\lambda_{14}| (r_4^2 + r_3 r_4)
 \nonumber\\
 &- \frac{1}{2}\, |\lambda_{15}| (r_1 r_3 + r_1 r_4)
 - \frac{1}{2}\, |\lambda_{16}| (r_2 r_3 + r_2 r_4) .
\end{align}
Putting everything together, the sufficient conditions for the quadratic potential
to be BFB are equivalent to requiring that the quadratic form
\begin{equation}
  \frac{1}{2} \sum_{ij} r_i \overline{A}_{ij} r_j, \quad \text{where}\quad
  \overline{A}=\begin{bmatrix}
    \bar{\lambda}_{11} &\bar{\lambda}_{12} &
    \bar{\lambda}_{13} &\bar{\lambda}_{14} \\
    \bar{\lambda}_{12} &\bar{\lambda}_{22} &
    \bar{\lambda}_{23} &\bar{\lambda}_{24} \\
    \bar{\lambda}_{13} &\bar{\lambda}_{23} &
    \bar{\lambda}_{33} &\bar{\lambda}_{34} \\
    \bar{\lambda}_{14} &\bar{\lambda}_{24} &
    \bar{\lambda}_{34} &\bar{\lambda}_{44} 
  \end{bmatrix},
\end{equation}
and
\begin{align}
  \bar{\lambda}_{11}=& \lambda_1\, ,
\nonumber\\
  \bar{\lambda}_{12}=& \lambda_3+\lambda_4+\text{min}(0,-\lambda_4)
  -|\lambda_5|\, ,
\nonumber\\ 
  \bar{\lambda}_{13}=& \frac{1}{2}\lambda_7
  -  \frac{1}{2} |\lambda_{15}|\, ,
\nonumber\\
  \bar{\lambda}_{14}=& \frac{1}{2}\lambda_{11}
  -  \frac{1}{2} |\lambda_{15}|\, ,
\nonumber\\
  \bar{\lambda}_{22}=& \lambda_2\nonumber\\
  \bar{\lambda}_{23}=& \frac{1}{2}\lambda_8
  -  \frac{1}{2} |\lambda_{16}|\, ,
\nonumber\\
  \bar{\lambda}_{24}=& \frac{1}{2}\lambda_{12}
  -  \frac{1}{2} |\lambda_{16}|\, ,
\nonumber\\
  \bar{\lambda}_{33}=& \frac{1}{4}\lambda_6
-  \frac{1}{3} |\lambda_{13}|\, ,
\nonumber\\
  \bar{\lambda}_{34}=& \frac{1}{4}\lambda_{10}
  -\frac{1}{6} |\lambda_{13}|
-   \frac{1}{6} |\lambda_{14}|\, ,
\nonumber\\
  \bar{\lambda}_{44}=& \frac{1}{4}\lambda_9 -
  \frac{1}{3} |\lambda_{14}|\, ,
\end{align}
is positive definite.
The criteria for verifying the copositivity of the $4\times 4$ matrix
$\bar{A}$ are given in Ref.~\cite{PING1993109}. We have implemented
this long algorithm (it has 49 steps) and verified, using the
\texttt{CERN} library
\texttt{Minuit}~\cite{james:1975dr}, that all points that passed the
above criteria were indeed BFB.

\section{\label{Global minimum}Global minimum}

It is not only necessary to ensure that the potential is bounded from
below but also to ensure that the minimum that we want to study is
indeed the global minimum. This is a more demanding task because there are
many other minima, and for some of them it is not possible to express analytically
the value of the potential at the minimum in terms of the parameters
of the potential. Just to fix notation we consider, in general,
\begin{equation}
  \label{eq:3}
  \vev{\Phi_1}=
  \begin{pmatrix}
    0\\
    \frac{1}{\sqrt{2}}v_1
  \end{pmatrix},\
  \vev{\Phi_2}=
  \begin{pmatrix}
    0\\
    \frac{1}{\sqrt{2}}v_2
  \end{pmatrix},\
  \vev{S}= v_S,\
  \vev{P}= v_P \, .
\end{equation}
As an example of a difficult situation is the case, $v_1=v_2=0$. The
minimization conditions lead to a system of coupled non-linear
equations with no simple analytical solution. However there are a few
that are simple to solve and we will give here their expressions.

\subsection{\texorpdfstring{$v_1\not=0, v_2\not=0$ and $v_S=v_P=0$}{v1 not 0, v2 not 0 and vs=vp=0}}

This is the case that we want to study. It is therefore important to
have an expression for this case so that we can compare with other
minima. We have
\begin{equation}
  \label{eq:4}
  V_N=\frac{1}{8} \left(-\lambda_1 v_1^4 - v_2^2 (2 \lambda_3 v_1^2 +
    2 \lambda_4 v_1^2  + 2 \lambda_5 v_1^2 + \lambda_2 v_2^2)\right)  \, .
\end{equation}

\subsection{\texorpdfstring{$v_1=0, v_2=0,v_S=0$ and $v_P\not=0$}{v1=v2=vS=0 and vP not 0}}

We have from the minimization equation\footnote{We do not check if it is a
minimum or a saddle point. This is because, if it is a saddle point,
there will be a minimum below that, and therefore the point should be
discarded.}
\begin{equation}
  \label{eq:6}
  v_P^2 = - \frac{2 m_P^2}{\lambda_9} \, .
\end{equation}
Then, if $v_P^2 > 0$, the potential at he the minimum is given by,
\begin{equation}
  \label{eq:7}
  V_{p}=- \frac{ m_P^2}{2 \lambda_9}  \, .
\end{equation}

\subsection{\texorpdfstring{$v_1=0, v_2=0,v_P=0$ and $v_S\not=0$}{v1=v2=vp=0 and vs not 0}}

We have from the minimization equation
\begin{equation}
  \label{eq:8}
  v_S^2 = - \frac{2 m_S^2}{\lambda_6} \, .
\end{equation}
Then, if $v_S^2 > 0$, the potential at the minimum is given by,
\begin{equation}
  \label{eq:9}
  V_{s}=- \frac{ m_S^2}{2 \lambda_6}  \, .
\end{equation}

\subsection{\texorpdfstring{$v_1=0, v_S=0$ and $v_2\not=0,v_P\not=0$}{v1=0, vs=0 and v2 not 0, vp not 0}}

We have from the minimization equation
\begin{align}
  \label{eq:10}
  v_P^2 =& - \frac{2 (2 \lambda_{12} m_{22}^2
    - \lambda_2 m_P^2)}{\lambda_{12}^2 - \lambda_2 \lambda_9} \, ,
  \nonumber\\
  v_2^2=&\frac{2 (2 \lambda_9 m_{22}^2 -
    \lambda_{12} m_P^2)}{\lambda_{12}^2 - \lambda_2 \lambda_9}  \, .
\end{align}
Then, if $v_P^2 > 0$ and $v_2^2>0$, the potential at the minimum
is given by, 
\begin{equation}
  \label{eq:11}
  V_{v_2 p}=\frac{4 \lambda_9 m_{22}^4 - 4 \lambda_{12} m_{22}^2 m_P^2 +
    \lambda_2 m_P^4}{2  \lambda_{12}^2 - 2 \lambda_2 \lambda_9}  \, .
\end{equation}

\subsection{\texorpdfstring{$v_1=0, v_P=0$ and $v_2\not=0,v_S\not=0$}{v1 = 0, v = 0 and v2 not 0,vS not 0}}

We have from the minimization equation
\begin{align}
  \label{eq:12}
  v_S^2 =&-\frac{2 (-2 \lambda_8 m_{22}^2 + \lambda_2
    m_S^2)}{\lambda_2 \lambda_6 - \lambda_8^2}  \, ,
  \nonumber\\
  v_2^2=&\frac{2 (-2 \lambda_6 m_{22}^2 + \lambda_8 m_S^2)}{\lambda_2
    \lambda_6 - \lambda_8^2}  \, .
\end{align}
Then, if $v_S^2 > 0$ and $v_2^2>0$, the potential at the minimum
is given by, 
\begin{equation}
  \label{eq:13}
  V_{v_2 s}=-\frac{4 \lambda_6 m_{22}^4 - 4 \lambda_8 m_{22}^2 m_S^2 +
    \lambda_2 m_S^4}{2   \lambda_2 \lambda_6 - 2 \lambda_8^2} \, .
\end{equation}

\subsection{\texorpdfstring{$v_2=0, v_S=0$ and $v_1\not=0,v_P\not=0$}{v2 = 0, vs = 0 and v1 not 0,vp not 0}}

We have from the minimization equation
\begin{align}
  \label{eq:14}
  v_P^2 =& -\frac{2 (2 \lambda_{11} m_{11}^2 - \lambda_1
  m_P^2)}{\lambda_{11}^2 - \lambda_1 \lambda_9}  \, ,
  \nonumber\\
  v_1^2=&\frac{2 (-2 \lambda_9 m_{11}^2 + \lambda_{11}
  m_P^2)}{-\lambda_{11}^2 + \lambda_1 \lambda_9 }  \, .
\end{align}
Then, if $v_P^2 > 0$ and $v_1^2>0$, the potential at the minimum
is given by, 
\begin{equation}
  \label{eq:15}
  V_{v_1 p}=\frac{4 \lambda_9 m_{11}^4 - 4 \lambda_ {11} m_{11}^2 m_P^2 +
  \lambda_1 m_P^4}{2 \lambda_{11}^2 - 2 \lambda_1 \lambda_9 }  \, .
\end{equation}

\subsection{\texorpdfstring{$v_2=0, v_P=0$ and $v_1\not=0,v_S\not=0$}{v2=0, vp=0 and v1 not 0, vs not 0}}

We have from the minimization equation
\begin{align}
  \label{eq:16}
  v_S^2 =& -\frac{2 (-2 \lambda_7 m_{11}^2 + \lambda_1 m_S^2)}{\lambda_1
  \lambda_6 - \lambda_7^2 }  \, ,
  \nonumber\\
  v_1^2=&\frac{2 (-2 \lambda_6 m_{11}^2 + \lambda_7 m_S^2)}{\lambda_1
  \lambda_6 - \lambda_7^2 }  \, .
\end{align}
Then, if $v_S^2 > 0$ and $v_1^2>0$, the potential at the minimum
is given by, 
\begin{equation}
  \label{eq:17}
  V_{v_1 s}=-\frac{4 \lambda_6 m_{11}^4 - 4 \lambda_7 m_{11}^2 m_S^2 +
  \lambda_1 m_S^4}{2 \lambda_1 \lambda_6 - 2 \lambda_7^2}  \, .
\end{equation}

\subsection{\texorpdfstring{$v_2=0, v_P=0, v_S=0$ and $v_1\not=0$}{v2=, vp=0, vs=0 and v1 not 0}}
We have from the minimization equation
\begin{align}
  \label{eq:18}
  v_1^2=&-\frac{4 m_{11}^2}{\lambda_1}  \, .
\end{align}
Then, if $v_1^2>0$, the potential at the minimum
is given by, 
\begin{equation}
  \label{eq:19}
  V_{v_1}=-\frac{2 m_{11}^4}{\lambda_1}  \, .
\end{equation}

\subsection{\texorpdfstring{$v_1=0, v_P=0, v_S=0$ and $v_2\not=0$}{v1=0, vP=0, vS=0 and v2 not 0}}

We have from the minimization equation
\begin{align}
  \label{eq:20}
  v_2^2=&-\frac{4 m_{22}^2}{\lambda_2}  \, .
\end{align}
Then, if $v_2^2>0$, the potential at the minimum
is given by, 
\begin{equation}
  \label{eq:21}
  V_{v_2}=-\frac{2 m_{22}^4}{\lambda_2} \, .
\end{equation}

\subsection{Other minima}

So, if our minimum, in eq.~(\ref{eq:4}), is above one of the cases in
eqs.~(\ref{eq:7}),(\ref{eq:9}), (\ref{eq:11}), (\ref{eq:13}),
(\ref{eq:15}), (\ref{eq:17}), (\ref{eq:19}),(\ref{eq:21}) then the
point in parameter space must be discarded. However, the list above
does not contain all the possible minima. We have verified with  a 
minimization procedure using \texttt{Minuit}~\cite{james:1975dr} that
there are other two cases. These do not have an analytical solution
(system of coupling cubic non-linear equations), but we have found a
way of solving them numerically. We will discuss these in the next two
subsections.

\subsection{\texorpdfstring{$v_S=v_P=0$, $v_1,v_2\not=0$, and $\alpha_2=\pi$}{vS=vP=0, v1,v2 not 0 and alp2=pi}}

The potential reads

\begin{equation}
  \label{eq:42}
  V_{\alpha_2=\pi}=\frac{1}{2} \left(m_{11}^2
   v_1^2+2 m_{12}^2 v_1 v_2+ m_{22}^2
   v_2^2\right) +
  \frac{1}{24} \left(3 \lambda_1 v_1^4+3 \lambda_2 v_2^4+6
   \lambda_{345} v_1^2 v_2^2\right) \, ,
\end{equation}
where
\begin{equation}
  \label{eq:49}
  \lambda_{345}=\lambda_3+\lambda_4+\lambda_5 \, .
\end{equation}
This leads to the minimization equations
\begin{align}
  2 \frac{\partial V}{\partial v_1}=&
 \, 2 m_{11}^2 v_1 + \lambda_1 v_1^3 + v_2 (2 m_{12}^2 + \lambda_{345}
  v_1 v_2)=0  \, ,
  \\[+2mm]
  2  \frac{\partial V}{\partial v_2}=&
 \,   2 m_{12}^2 v_1 + 2 m_{22}^2 v_2 + \lambda_{345} v_1^2 v_2 +
  \lambda_2 v_2^3=0  \, .
\end{align}
This is a system of cubic non-linear equations with no analytical
solution. However we devised a way of solving it numerically. The idea
is to define
\begin{equation}
  \label{eq:50}
  v_1= v \cos\beta, \ v_2=v \sin\beta  \, ,
\end{equation}
where $v,\beta$ should not be confused with the quantities of the same
name for our minimum. Substituting and rearranging we get two equations,
\begin{align}
0=&  \,  2 \lambda_1 m_{12}^2 \cos^2\beta - 2 \lambda_{345} m_{11}^2
\cos\beta \sin\beta +  
 2 \lambda_1 m_{22}^2 \cos\beta \sin\beta - 2 \lambda_2 m_{11}^2
 \sin^2\beta \tan\beta\nonumber\\
& +  2 \lambda_{345} m_{22}^2 \sin^2\beta \tan\beta - 
 2 \lambda_2 m_{12}^2 \sin^2\beta \tan^2\beta  \, ,\\[+2mm]
 v^2=&  \,  \frac{-2 m_{11}^2 - 2 m_{12}^2 \tan\beta}{
 \lambda_1 \cos^2\beta + \lambda_{345} \sin^2\beta}  \, .
\end{align}
Now we know that with our choices $\beta \in [0,\pi/2]$. So we solve
numerically 
the first equation for $\beta$ in that interval. If it has a solution,
we substitute it in the second equation. If $v^2 > 0$ we substitute in
eq.~(\ref{eq:50}) and then in eq.~(\ref{eq:42}). We finally
compare this value with the value of the inert minimum,
eq.~(\ref{eq:4}). If it is lower, we discard the point.

\subsection{\texorpdfstring{$v_1=v_2=0$, $v_S,v_P\not=0$}{v1=v2=0, vs,vp not 0}}

Finally we consider the last case, $v_1=v_2=0$, $v_S,v_P\not=0$. We
adopt a strategy similar to that of the last case. First the potential
reduces to
\begin{align}
  \label{eq:51}
V_{s,p}=&  \frac{1}{2}\left(m_P^2\ v_{P}^2-2 m_{SP}\  v_{P}\, v_{S}+
  m_S^2\ v_{S}^2\right) 
 \nonumber\\
&+ 
  \frac{1}{24} \left(6 \lambda_{10}\ v_{P}^2\, v_{S}^2+4 \lambda_{13}\ v_{P}\,
    v_{S}^3+4 \lambda_{14}\ 
    v_{P}^3\, v_{S}+3 \lambda_6\ v_{S}^4+3 \lambda_9\ v_{P}^4\right)\ .
\end{align}
The stationary equations read
\begin{align}
  \label{eq:53}
   \frac{\partial V}{\partial v_{S}}=&
  \frac{1}{2} \lambda_{10}\, v_{P}^2 v_{S}+\frac{1}{2} \lambda_{13}\, v_{P}\,
  v_{S}^2+\frac{\lambda_{14}\, v_{P}^3}{6}+\frac{\lambda_6\, v_{S}^3}{2}
  -m_{SP}^2\, v_{P}+ m_S^2\, v_{S} =0\ ,
  \\[+2mm]
    \frac{\partial V}{\partial v_{P}}=&
  \frac{1}{2} \lambda_{10}\, v_{P}\, v_{S}^2+\frac{\lambda_{13}\,
    v_{S}^3}{6}+\frac{1}{2} 
  \lambda_{14}\, v_{P}^2 v_{S}+\frac{\lambda_9\, v_{P}^3}{2}+m_P^2\, v_{P}
  -m_{SP}^2\, v_{S} = 0\ .
\end{align}
Now we define
\begin{equation}
  \label{eq:52}
   v_{P} = v_{S} \, \eta \ ,
\end{equation}
and then eq.~(\ref{eq:53}) can be solved for the following two
equations,
\begin{align}
  \label{eq:54}
0= & \frac{1}{6} \eta^4 \lambda_{14} m_P^2+\frac{1}{2} \eta^4
   \lambda_9 m_{SP}^2+\frac{1}{2} \eta^3 \lambda_{10}
   m_P^2+\frac{1}{3} \eta^3 \lambda_{14} m_{SP}^2-\frac{1}{2}
   \eta^3 \lambda_9 m_S^2+\frac{1}{2} \eta^2 \lambda_{13}
   m_P^2\nonumber\\
&   -\frac{1}{2} \eta^2 \lambda_{14} m_S^2-\frac{\eta 
   \lambda_{10} m_S^2}{2}-\frac{\eta  \lambda_{13}
   m_{SP}^2}{3}+\frac{\eta  \lambda_6
   m_P^2}{2}-\frac{\lambda_{13} m_S^2}{6}-\frac{\lambda_6
   m_{SP}^2}{2} \ ,
\end{align}
for $\eta$ and
\begin{align}
  \label{eq:55}
  v_{S}^2=\frac{6 \eta\,  m_{SP}^2-6 m_S^2}{\eta ^3\, \lambda_{14}+3 \eta^2\,
   \lambda_{10}+3 \eta\,  \lambda_{13}+3 \lambda_6}\ ,
\end{align}
for $v_{S}^2$. The procedure follows as before. We solve numerically
eq.~(\ref{eq:54}). Then substitute in eq.~(\ref{eq:55}). If $v_{S}^2 > 0$, then 
we get the value of the potential at this stationary point
substituting back in eq.~(\ref{eq:53}), after using
eq.~(\ref{eq:52}). Then we compare with our minimum, eq.~(\ref{eq:4}),
and discard the point if this mininum is lower than our desired
minimum. There is only one subtle point here. In principle $v_S,v_P$
can have either sign, which means that also $\eta$ can be positive or
negative. So we have to find solutions for $\eta$ in both cases. Then
we proceed to check if $v_S^2>0$. If this is the case we still have
the posibility of $v_S=\pm \sqrt{v_S^2}$. So we have, in principle,
four possibilities, two signs for $\eta$ and two signs for $v_S$. We
calculate the value of the potential in eq.~(\ref{eq:51}) for all possibilities, and compare
with our desired minimum,  eq.~(\ref{eq:4}).

\section{Perturbative Unitarity}
\label{sec:Unitarity}

For the perturbative unitarity, we follow section 4 of
Ref.~\cite{Bento:2017eti}. To be self-contained we reproduce here their
table 3 for our case. For convenience we define
\begin{equation}
  \label{eq:36}
  s_1=s, \quad s_2=p \, ,
\end{equation}
and we get the results in table~\ref{tab:unitarity1}. Notice that in
comparison with Table 3 of Ref.~\cite{Bento:2017eti}, we have
\begin{table}[h!]
  \centering
  \begin{tabular}{|c|c|l|c|}\hline
    $Q$&$2\mathcal{Y}$&State&\# states\\[+1mm]\hline\hline
    2 &2 &$S_{\alpha}^{++}=\{w_1^+ w_1^+,w_1^+ w_2^+,w_2^+ w_2^+\}$&$3$
    \\[+1mm]\hline\hline
    1 &2 & $S_{\alpha}^{+}=\{w_1^+ n_1,w_1^+ n_2,w_2^+ n_1,w_2^+n_2\}$&$4$
    \\[+1mm]\hline
    1 &1 & $R_{\alpha}^{+}=\{w_1^+ s_1,w_1^+ s_2,w_2^+ s_1,w_2^+s_2\}$&$4$
    \\[+1mm]\hline
1 &0 & $T_{\alpha}^{+}=\{w_1^+ n^*_1,w_1^+ n^*_2,w_2^+ n^*_1,w_2^+ n^*_2\}$
& $4$\\[+1mm]\hline\hline
0 &2 & $S_{\alpha}^{0}=\{n_1 n_1,n_1 n_2,n_2 n_2\}$
& $3$ \\[+1mm]\hline
0 &1 & $P_{\alpha}^{0}=\{n_1 s_1,n_1 s_2,n_2 s_1,n_2 s_2\}$
& $4$ \\[+1mm]\hline
0 &0 &$T_{\alpha}^{0}= \{w_1^- w_1^+,w_1^- w_2^+,w_2^- w_1^+,w_2^- w_2^+,n_1
n^*_1,n_1 n^*_2,n_2 n^*_1,n_2 n^*_2,s_1 s_1, s_1 s_2, s_2 s_2\}$
&$11$  \\[+1mm]\hline
  \end{tabular}
  \caption{List of two body scalar states separated by $(Q,\mathcal{Y})$.}
  \label{tab:unitarity1}
\end{table}
two more possibilities, that we denote by $R^+_\alpha$ with charge 1
and hypercharge 1, and $P_\alpha^0$ with charge 0 and hypercharge
1. These are not present in NHDM. We have therefore to obtain seven
matrices. From these four are equal to the 2HDM one is different and
two are new.

Applying the procedure described in Ref.~\cite{Bento:2017eti} we get
the following results with the notation $M_{\mathcal{Y}}^{\mathcal{Q}}$.
\begin{equation}
  \label{eq:37}
  M_2^{++}=
  \begin{bmatrix}
    \lambda_1 &0 & \lambda_5\\
    0&\lambda_3+\lambda_4&0\\
    \lambda_5&0&\lambda_2
  \end{bmatrix} \, ,
\end{equation}
with eigenvalues
\begin{equation}
  \label{eq:38}
  \lambda_3 + \lambda_4,\ \frac{1}{2} (\lambda_1 + \lambda_2 \pm
  \sqrt{\lambda_1^2 - 2 \lambda_1 \lambda_2 + \lambda_2^2 + 4
    \lambda_5^2}) \, .
\end{equation}
\begin{equation}
  \label{eq:39}
  M_2^+=
  \begin{bmatrix}
    \lambda_1& 0& 0& \lambda_5\\
    0& \lambda_3& \lambda_4& 0\\
    0& \lambda_4& \lambda_3& 0\\
    \lambda_5& 0& 0& \lambda_2
  \end{bmatrix} \, ,
\end{equation}
with eigenvalues
\begin{equation}
  \label{eq:40}
  \lambda_3 - \lambda_4, \lambda_3 + \lambda_4, 
 \frac{1}{2} \left(\lambda_1 + \lambda_2 \pm \sqrt{\lambda_1^2 - 2
     \lambda_1 \lambda_2 + \lambda_2^2 + 4 \lambda_5^2}\right) \, .
\end{equation}
\begin{equation}
  \label{eq:41}
  M_1^+=
  \begin{bmatrix}
    \lambda_7& \frac{1}{2} \lambda_{15}& 0& 0\\
    \frac{1}{2}\lambda_{15}& \lambda_{11}& 0& 0\\
    0& 0& \lambda_8& \frac{1}{2} \lambda_{16}\\
    0& 0 &\frac{1}{2}\lambda_{16}& \lambda_{12}
  \end{bmatrix} \, ,
\end{equation}
with eigenvalues
\begin{equation}
  \label{eq:43}
  \frac{1}{2} (\lambda_{11} + \lambda_7 \pm \sqrt{\lambda_{11}^2 +
    \lambda_{15}^2 - 2 \lambda_{11} \lambda_7 + \lambda_7^2}),  
 \frac{1}{2} (\lambda_{12} + \lambda_8 \pm \sqrt{\lambda_{12}^2 +
   \lambda_{16}^2 - 2 \lambda_{12} \lambda_8 + \lambda_8^2})  \, .
\end{equation}
\begin{equation}
  \label{eq:44}
  M_0^+=
  \begin{bmatrix}
   \lambda_1& 0& 0& \lambda_4\\
   0 & \lambda_3& \lambda_5& 0\\
   0 & \lambda_5& \lambda_3& 0\\
   \lambda_4& 0& 0& \lambda_2\\
  \end{bmatrix} \, ,
\end{equation}
with eigenvalues,
\begin{equation}
  \label{eq:45}
 \lambda_3 - \lambda_5, 
 \lambda_3 + \lambda_5,  \frac{1}{2} (\lambda_1 + \lambda_2 \pm
 \sqrt{\lambda_1^2 - 2 \lambda_1 \lambda_2 + \lambda_2^2 + 4  \lambda_4^2}) \, .
 \end{equation}
 \begin{equation}
   \label{eq:46}
   M_2^0= M_2^{++} \, ,
 \end{equation}
 with the same eigenvalues as in eq.~(\ref{eq:38}).
 \begin{equation}
   \label{eq:47}
   M_1^0=M_1^+ \, ,
 \end{equation}
 with the same eigenvalues as in eq.~(\ref{eq:43}). Finally
 \begin{equation}
   \label{eq:48}
   M_0^0=
  \left[
\begin{array}{ccccccccccc}
  2 \lambda_1& 0& 0& \lambda_3 + \lambda_4& \lambda_1& 0& 0&  \lambda_3& \frac{\lambda_7}{\sqrt{2}}& \frac{\lambda_{15}}{2}&
  \frac{\lambda_{11}}{\sqrt{2}}\\  
  0& \lambda_3 + \lambda_4& 2 \lambda_5& 0& 0& \lambda_4& \lambda_5&    0& 0& 0& 0\\ 
  0& 2 \lambda_5&  \lambda_3 + \lambda_4& 0& 0& \lambda_5& \lambda_4&
  0& 0& 0& 0\\ 
  \lambda_3 + \lambda_4& 0& 0& 2 \lambda_2& \lambda_3& 0& 0&
  \lambda_2& \frac{\lambda_8}{\sqrt{2}}& \frac{\lambda_{16}}{2}&
  \frac{\lambda_{12}}{\sqrt{2}}\\
  \lambda_1& 0& 0& \lambda_3& 2 \lambda_1& 0& 0& \lambda_3 +
  \lambda_4& \frac{\lambda_7}{\sqrt{2}}& \frac{\lambda_{15}}{2}&
  \frac{\lambda_{11}}{\sqrt{2}}\\  
  0& \lambda_4& \lambda_5& 0& 0& \lambda_3 + \lambda_4&2 \lambda_5&
    0& 0& 0& 0\\ 
  0& \lambda_5& \lambda_4& 0& 0& 2 \lambda_5& \lambda_3 + \lambda_4&
  0& 0& 0& 0\\ 
  \lambda_3& 0& 0& \lambda_2& \lambda_3 + \lambda_4& 0& 0& 2
  \lambda_2& \frac{\lambda_8}{\sqrt{2}}& \frac{\lambda_{16}}{2}&
  \frac{\lambda_{12}}{\sqrt{2}}\\  
  \frac{\lambda_7}{\sqrt{2}}& 0& 0& \frac{\lambda_8}{\sqrt{2}}&
  \frac{\lambda_7}{\sqrt{2}}& 0& 
  0& \frac{\lambda_8}{\sqrt{2}}& \frac{3\lambda_6}{2}&
  \frac{\lambda_{13}}{\sqrt{2}}& 
  \frac{\lambda_{10}}{2}\\ 
  \frac{\lambda_{15}}{2}& 0& 0& \frac{\lambda_{16}}{2}&
  \frac{\lambda_{15}}{2}& 0& 0& 
  \frac{\lambda_{16}}{2}& \frac{\lambda_{13}}{\sqrt{2}}& \lambda_{10}& 
    \frac{\lambda_{14}}{\sqrt{2}}\\
  \frac{\lambda_{11}}{\sqrt{2}}& 0& 0& \frac{\lambda_{12}}{\sqrt{2}}&
    \frac{\lambda_{11}}{\sqrt{2}}& 0& 0& \frac{\lambda_{12}}{\sqrt{2}}& 
    \frac{\lambda_{10}}{2}& \frac{\lambda_{14}}{\sqrt{2}}& \frac{3\lambda_9}{2} 
\end{array}
\right] .
 \end{equation}
 The eigenvalues separate in six that can be easily evaluated, but the
 other five are solutions of a fifth order polynomial. In this case it
 is easier to use the methods of the minors as explained in
 Ref.~\cite{Bento:2022vsb} for the whole matrix.

\section{\texorpdfstring{The precision observables $S,T$ and $U$}{The precision observables S,T and U}}
\label{sec:STU}

To obtain these observables one follows the setup of
Ref.~\cite{Grimus:2007if}. For this one has to obtain the matrices $U$
and $V$ in their notation. 

Following Ref.~\cite{Grimus:2007if}, and using the definitions given in section \ref{model} we define
\begin{equation}
  \label{eq:27}
  \begin{bmatrix}
    \varphi_1^0\\
    \varphi_2^0
  \end{bmatrix}
  =
  \begin{bmatrix}
    y_1 + i z_1\\
    y_2 + i z_2
  \end{bmatrix},\qquad
  \begin{bmatrix}
    \chi_1^0\\
    \chi_2^0
  \end{bmatrix}
  =
  \begin{bmatrix}
    s\\
    p
  \end{bmatrix} .
\end{equation}

We have now everything to define the matrices $U_{2\times 2}$,
$V_{2\times 6}$ and $R_{2 \times 6}.$\footnote{As can be seen from the
  expressions  for $S,T,U$ in Ref.~\cite{Grimus:2007if},
the matrix $R$  is not necessary.}
We get,
\begin{equation}
  \label{eq:28}
  U= \mathcal{O}_\beta^T =
  \begin{bmatrix}
    c_\beta & -s_\beta\\
    s_\beta & c_\beta 
  \end{bmatrix} ,
\end{equation}
\begin{equation}
  \label{eq:29}
  V=
  \begin{bmatrix}
    i\, c_\beta & c_{\alpha_1}& s_{\alpha_1}&-i\, s_\beta& 0& 0\\
    i\,s_\beta &-s_{\alpha_1}&c_{\alpha_1}&i\, c_\beta&0&0
  \end{bmatrix},
\end{equation}
\begin{equation}
  \label{eq:30}
  R=
  \begin{bmatrix}
    0&0&0&0&c_{\alpha_2}&s_{\alpha_2}\\
    0&0&0&0&-s_{\alpha_2}&c_{\alpha_2}
  \end{bmatrix},
\end{equation}
where we have organized the physical fields in the order
\begin{equation}
  \label{eq:31}
  (G_0,h_1,h_2,A,h_3,h_4)^T \, .
\end{equation}
The matrices required for the calculation of the precision
observables are,
\begin{equation}
  \label{eq:32}
  U^\dagger U =
  \begin{bmatrix}
    1 &0\\
    0&1
  \end{bmatrix},
\end{equation}
\begin{equation}
  \label{eq:33}
  U^\dagger V =
  \left[
\begin{array}{cccccc}
 i & c_{\alpha_1} c_\beta-s_{\alpha_1} s_\beta & c_{\alpha_1}
   s_\beta+c_\beta s_{\alpha_1} & 0 & 0 & 0 \\
 0 & -c_{\alpha_1} s_\beta-c_\beta s_{\alpha_1} & c_{\alpha_1}
   c_\beta-s_{\alpha_1} s_\beta & i & 0 & 0 \\
\end{array}
\right] \, ,
\end{equation}
\begin{equation}
  \label{eq:34}
  V^\dagger V=\left[
\begin{array}{cccccc}
 1 & i s_{\alpha_1} s_\beta-i c_{\alpha_1} c_\beta & -i c_{\alpha_1} s_\beta-i
   c_\beta s_{\alpha_1} & 0 & 0 & 0 \\
 i c_{\alpha_1} c_\beta-i s_{\alpha_1} s_\beta & 1 & 0 & -i c_{\alpha_1}
   s_\beta-i c_\beta s_{\alpha_1} & 0 & 0 \\
 i c_{\alpha_1} s_\beta+i c_\beta s_{\alpha_1} & 0 & 1 & i c_{\alpha_1}
   c_\beta-i s_{\alpha_1} s_\beta & 0 & 0 \\
 0 & i c_{\alpha_1} s_\beta+i c_\beta s_{\alpha_1} & i s_{\alpha_1} s_\beta-i
   c_{\alpha_1} c_\beta & 1 & 0 & 0 \\
 0 & 0 & 0 & 0 & 0 & 0 \\
 0 & 0 & 0 & 0 & 0 & 0 \\
\end{array}
\right]\, ,
\end{equation}
\begin{equation}
  \label{eq:35}
  \text{Im}(V^\dagger V)=
  \left[
\begin{array}{cccccc}
 0 & s_{\alpha_1} s_\beta-c_{\alpha_1} c_\beta & -c_{\alpha_1}
   s_\beta-c_\beta s_{\alpha_1} & 0 & 0 & 0 \\
 c_{\alpha_1} c_\beta-s_{\alpha_1} s_\beta & 0 & 0 & -c_{\alpha_1}
   s_\beta-c_\beta s_{\alpha_1} & 0 & 0 \\
 c_{\alpha_1} s_\beta+c_\beta s_{\alpha_1} & 0 & 0 & c_{\alpha_1}
   c_\beta-s_{\alpha_1} s_\beta & 0 & 0 \\
 0 & c_{\alpha_1} s_\beta+c_\beta s_{\alpha_1} & s_{\alpha_1}
   s_\beta-c_{\alpha_1} c_\beta & 0 & 0 & 0 \\
 0 & 0 & 0 & 0 & 0 & 0 \\
 0 & 0 & 0 & 0 & 0 & 0 \\
\end{array}
\right] .
\end{equation}
Using these matrices, we have programmed the expressions in Ref.~\cite{Grimus:2007if}. The only important point  to remember is the order
of the fields in eq.~(\ref{eq:31}).

 \section{Experimental Constraints}\label{sec:exp}

We have discussed the conditions for boundedness from below,
a global minimum and perturbative unitarity. We complete the
theoretical constraints by requiring perturbativity of the Type-II Yukawa couplings,
by setting them to be $|y_i| < \sqrt{4\pi}$, where $i = t, b, \tau$.
The bounds from experimental collider searches follow: the oblique parameters STU
described in Section~\ref{sec:STU} are to be compared with the global electroweak
fit in~\cite{Baak:2014ora}; the coupling-strength modifiers within $3\sigma$ of
the LHC data~\cite{ATLAS:2019nkf}; the LHC signal strengths of the $125\textrm{GeV}$ Higgs,
for the combinations of production cross sections and branching ratios,
to have a $2\sigma$ agreement with the
most recent ATLAS results~\cite{ATLAS:2022vkf}.
For LHC searches for new particles,
we use the software package \texttt{HiggsTools-1.1.3}~\cite{Bahl:2022igd},
which includes the latest data from the ATLAS and
CMS experiments at CERN, and the latest LHC bound on the invisible branching ratio
of the Higgs, BR($h\rightarrow \text{inv}$) $\leq$ 0.107 at $95\%$ C.L.~\cite{ATLAS:2023tkt}.

For the DM observables, we implemented our model in
\texttt{micrOMEGAs-6.2.3}~\cite{Alguero:2023zol} to numerically calculate the relic density,
scattering amplitudes, annihilation cross section and decay rates. We note that the $\Z2$ symmetry used in this work does \textit{not}
separate the DM particle into two non-communicating sectors. One
could think that a second DM particle could be invoked by imposing
a kinematic limit on the mass of the second DM candidate. However,
in this case, such putative DM candidate could decay into the
(true) lightest DM candidate and a SM pair, in particular through
the terms proportional to $\lambda_{15}$ and $\lambda_{16}$
in Eq.~\eqref{2hdm2s}.
Such decays are included using the \texttt{micrOMEGAs} routine
\texttt{darkOmegaN} but not under the \texttt{darkOmega2} routine~\cite{Alguero:2023zol}, which
comes activated by default in the usual \texttt{micrOMEGAs} distribution.
We compare with the Planck experiment~\cite{Planck:2018vyg}  at $3\sigma$,
\begin{equation}
    \Omega h^2= 0.1200\pm 0.0012\, .
\label{Planck}
\end{equation}
While optimizing our initial simulations and comparing methods,
we consider a more permissive range $0.09 < \Omega h^2 < 0.15$,
as identified appropriately in the respective captions. 

For the DM-nucleon scattering, we follow the
method described in Ref.~\cite{Belanger:2014bga} of computing the
normalized cross section of DM on a point-like Xenon nucleus
\begin{equation} \label{sigmascat}
\sigma_{\text{SI}}^{\text{Xe}}=
\frac{4 \mu_{\text{DM}}^2}{\pi}\frac{\left(Z f_p + (A-Z)f_n\right)^2}{A^2},
\end{equation}
with $\mu_{\text{DM}}$ the reduced mass of the DM candidate and
$f_p$, $f_n$ the amplitudes for protons and neutrons.
We compare with the most recent LZ release in 2024~\cite{LZ:2024zvo}. 

\begin{figure}[htb]
	\centering
	\includegraphics[height=6cm]{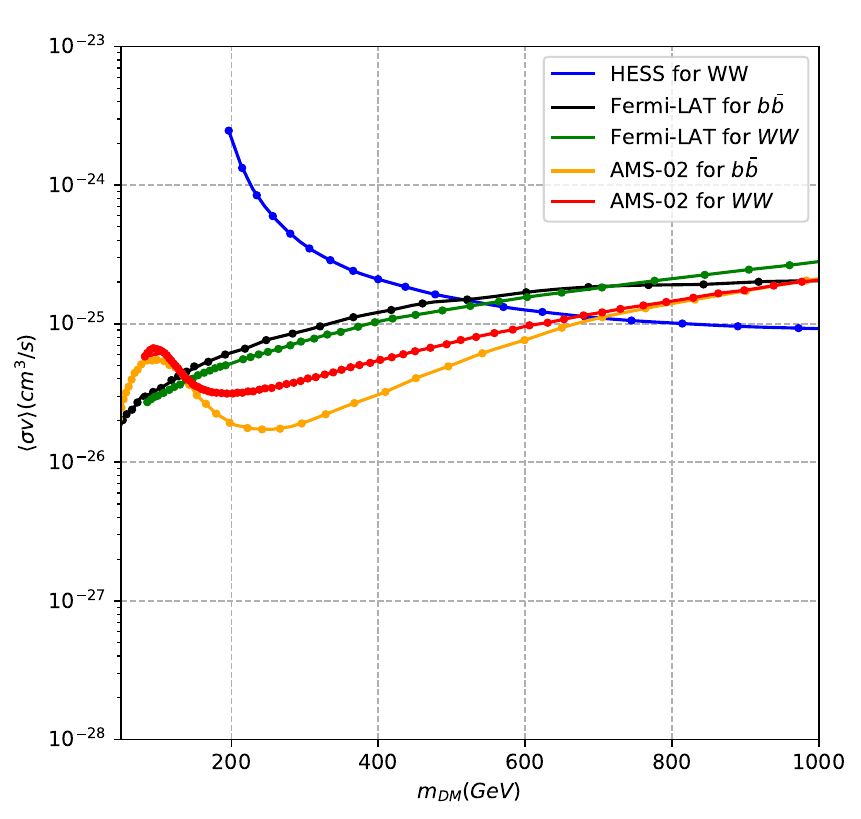}
\caption{\label{indirect_lines} Combined relevant limits from indirect
searches on the total $\langle\sigma v\rangle$ as a function of the
mass of the DM candidate $m_{DM}$.
The lines coming from Fermi-LAT~\cite{Fermi-LAT:2015att}
and H.E.S.S.~\cite{HESS:2022ygk} assume a Navarro-Frenk-White (NFW)
DM density profile and the AMS-02~\cite{AMS:2016oqu} lines
correspond to the conservative approach derived in Ref.~\cite{Reinert:2017aga}. }
\end{figure}

The annihilation cross section calculated is to be compared with
reconstructions based on indirect searches.
We follow the method described in~\cite{Belanger:2021lwd} and compare
the dominating channel with the respective experimental exclusion line.
We find that for our model and mass ranges, the annihilation occurs either
into $VV$, summing only the  $WW$ and $ZZ$ final states,
dubbed $\langle\sigma v\rangle_{VV}$, or into $b\bar{b}$
\footnote{There are currently no published equivalent exclusion lines for
decays going mainly into $c\bar{c}$ nor $h_1h_1$.}.
For each point in parameter space, the procedure is to check the
annihilation channel and apply the respective exclusion bounds in
Fig.~\ref{indirect_lines}.
We have confirmed that for the mass region studied, this constraint
does not exclude a significant amount of the parameter space when applied
after demanding the correct relic density.

\section{\label{sec:samp}Sampling Methods}

To explore the parameter space,
we consider three strategies:
i)\, a random scan without any prior assumptions on the parameter space;
ii)\, a scan close to the alignment limit of the 2HDM,
defined as $\alpha_1=-\beta$ for the choices in eqs.~(\ref{obeta}) and (\ref{field_eq});
iii)\, employing the Artificial Intelligence black box optimization approach first
presented in~\cite{deSouza:2022uhk},
applied to a real 3HDM in~\cite{Romao:2024gjx},
and to a complex 3HDM in~\cite{deSouza:2025bpl}.
Starting with random values for all parameters in a 2HDM with Type-II Yukawa couplings,
we quickly reproduce the lower bound on the mass of the only
charged Higgs boson which, at 95\% CL (2$\sigma$),
is according to~\cite{Misiak:2017bgg}:
\begin{equation}
  \label{eq:mc}
  m_{H^+}> 580\, \text{GeV}\, .
\end{equation}
We continued with a longer duration scan of the parameter space.
Our fixed inputs are $v = 246\,\text{GeV}$ and $m_{h1} = 125\,\text{GeV}$.
We then took random values in the ranges:
 \begin{subequations}
\label{eq:scanparameters}
	\begin{eqnarray}
&\alpha_1,\, \alpha_2 \in \left[-\frac{\pi}{2},\frac{\pi}{2}\right];\qquad \tan{\beta}\,\in \left[0.3,10\right];\\[8pt]
&m_{h2},\, m_{h3},\,m_{h4}\,
\in \left[125,1000\right]\,\text{GeV};\\[8pt]
&
m_{A},\,
\in \left[100,1000\right]\,\text{GeV};\qquad m_{H^\pm}\in \left[580,1000\right]\,\text{GeV};\\[8pt]
&
m^2_{12},m^2_{S},m^2_{P},m^2_{SP} \in  \left[\pm 10^{-1},\pm 10^{7}\right]\,
\text{GeV}^2\, ;\\[8pt]
&
\lambda_{6}, \lambda_{8}, \lambda_{9}, \lambda_{10}, \lambda_{12}, \lambda_{13}, \lambda_{14}, \lambda_{16}\in  \left[\pm 10^{-3},\pm 10^{1}\right]\,.
	\end{eqnarray}
\end{subequations}
For the sampling near the alignment limit,
we consider the same range, with the change that $\alpha_1$ is obtained as a random number
within $\pm 10\%$ of  $-\beta$.
We let each method run for about $\sim 7000$ CPU hours
in order to obtain $\sim 120000$ points for the random scan
and $\sim 370000$ points with the near alignment consideration.

\begin{figure}[htb]
	\centering
	\includegraphics[height=6cm]{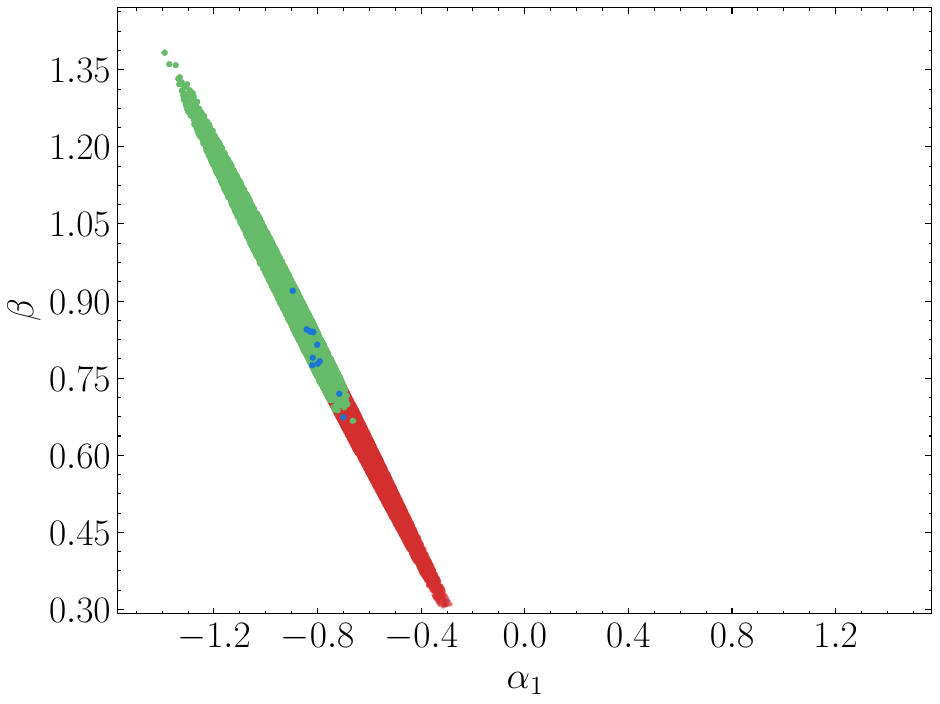}
\caption{\label{random_sample} Points obtained with a random sampling using
the ranges of parameters in eq.~\eqref{eq:scanparameters},
shown in the $\alpha_1-\beta$ plane.
The points in \textit{red} consider the expressions of Sections~\ref{vacuum}-\ref{sec:exp},
in order to satisfy BFB, unitarity, global minimum, flavour bounds,
coupling modifiers and signal strengths.
The points in \textit{green} combine points originally in red that are
found to also satisfy \texttt{HiggsTools-1.1.3}.
The \textit{blue} point is a green point that, in addition, meets the
condition $\Omega h^2 \in [0.09,0.15]$.}
\end{figure}

\begin{figure}[htb]
	\centering
	\includegraphics[height=6cm]{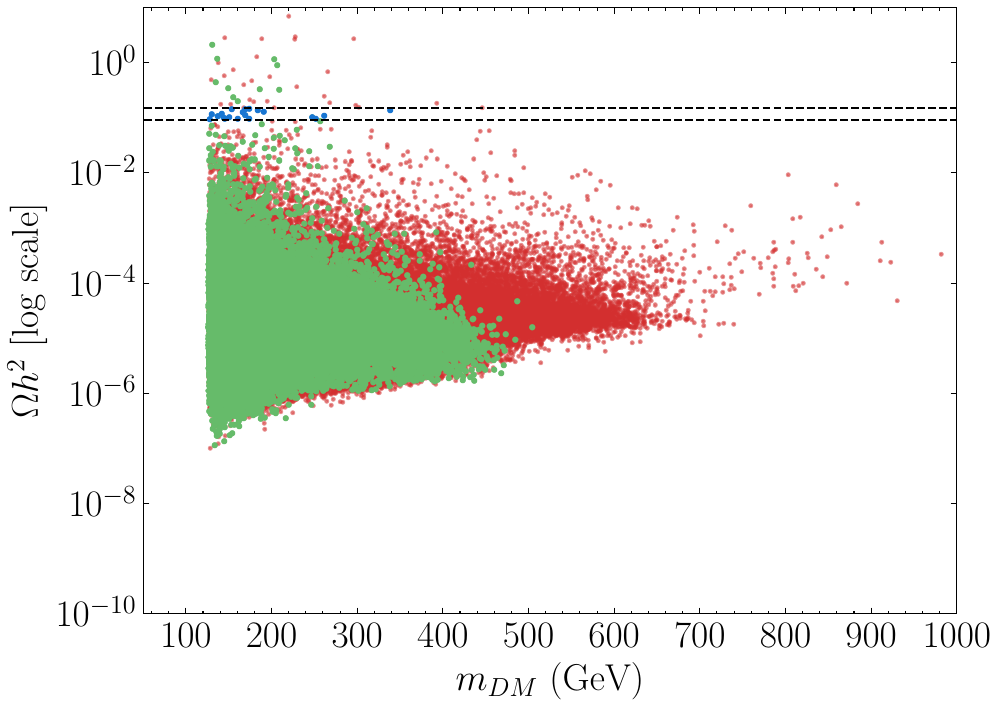}
\caption{\label{omega_random} Results in the mass - relic density plane for the scan
near the alignment limit,
considering eq.~\eqref{eq:scanparameters}, except for $\alpha$ obtained as a
random number within $\pm 10\%$ of  $-\beta$.
The color code coincides with Fig.~\ref{random_sample}. }
\end{figure}

We present the results of the random scan in Fig.~\ref{random_sample},
for the $\alpha_1-\beta$ plane,
showing clearly that considering $\alpha_1$ within $\pm 10\%$ of $\beta$ correctly
explores the allowed parameter space, and the scan near the alignment limit in Fig.~\ref{omega_random} for the mass - relic density plane.
The points in \textit{red} satisfy BFB, unitarity, global minimum,
flavour bounds, coupling modifiers and signal strengths.
The points in \textit{green} combine points originally in red that are found to
also satisfy \texttt{HiggsTools-1.1.3}.
We add in \textit{blue} the points originally in green found
to have a relic density of $\Omega h^2 \in [0.09,0.15]$.

\begin{figure}[htb]
	\centering
	\includegraphics[height=6cm]{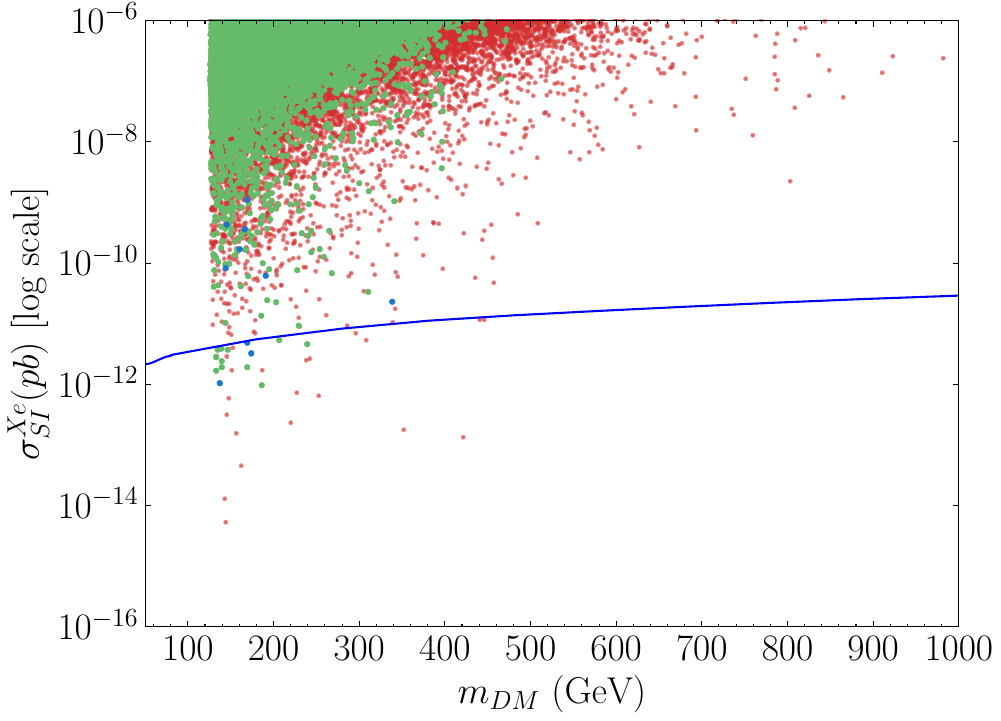}
\caption{\label{direct_k10} Direct detection results for the scan near
the alignment limit $\alpha_1=-[0.9,1.1]\beta$,
with the relevant quantity for nucleon scattering $\sigma_{SI}^{Xe}$
obtained from eq.~\eqref{sigmascat}.
The color code is the same as Fig.~\ref{random_sample},
with the addition of the LZ exclusion line from 2024~\cite{LZ:2024zvo}
drawn in blue.}
\end{figure}

At this stage, we do not impose the bounds from Dark Matter direct and indirect searches.
Such bounds rule out most of the few points we found with acceptable relic density,
with Fig.\ref{direct_k10} showing the comparison of the points sampled near the alignment
limit with the most recent exclusion line from the LZ collaboration~\cite{LZ:2024zvo}.
As most points sampled do not have a significant  relic density,
we continue with a new Machine Learning method instead of continuing with longer sets of
inefficient traditional sampling.

We followed the procedure  for efficiently sampling the parameter
space shown in \cite{deSouza:2022uhk} and further developed in
\cite{Romao:2024gjx,deSouza:2025uxb,deSouza:2025bpl} \footnote{Alternative approaches to successfully sample the parameter space of BSM models have been developed in cases where training datasets are available~\cite{Caron:2019xkx,Hollingsworth:2021sii,Goodsell:2022beo,Diaz:2024yfu,Batra:2025amk,Hammad:2025wst}. For more broad applications of Machine Learning in the field of particle physics, consider the comprehensive reviews~\cite{Feickert:2021ajf,Plehn:2022ftl}.}.
The first step is to define constraint functions, $C(\mathcal{O})$, as
\begin{equation}
    C(\mathcal{O}) = \max(0, -\mathcal{O} + \mathcal{O}_{LB}, \mathcal{O} - \mathcal{O}_{UB}),
\end{equation}
with $\mathcal{O}$ the value of a quantity constrained to be inside the interval
$[\mathcal{O}_{LB},\mathcal{O}_{UB}]$.
$C(\mathcal{O})$ then quantifies \textit{how far} the value of the observable is
from the defined bounds, or zero if within the specified interval.
The quantities $\mathcal{O}$ are obtained by a black box  computational routine that
takes a set of parameters $\theta$ as inputs to calculate all
relevant physical quantities $\mathcal{O}(\theta)$.
We follow the single-objective optimisation algorithm, obtaining the loss function as the sum of 
all the constraint functions of the model
\begin{equation} \label{eq:lossf}
    L(\theta) = \sum_{i=1}^{N_c} C(\mathcal{O}_i(\theta)),
\end{equation}
where the sum runs over all the $N_c$ constraints,
with $L = 0$ only when all constraints are
satisfied. 
One of the main strengths of the method is that the quantity
$\mathcal{O}_i$ does not need to be an experimental observable,
also allowing theoretical constraints and cuts in the same loss
function.

The optimization algorithm is the
Covariant Matrix Adaptation Evolutionary Strategy
(CMA-ES)~\cite{Hansen2001,hansen2023cmaevolutionstrategytutorial},
characterized by an iterative sampling according to a multivariate normal distribution,
initialized with its mean at a random point in parameter space and its
covariance matrix set to the identity matrix, $\one$, scaled by a constant.
A generation of candidate solutions is sampled from this distribution and 
the candidates are ranked from best to worst based on the loss function in
eq.~\eqref{eq:lossf}. The best candidates are then used to
compute a new mean and approximate the covariance matrix,
for the next iterative step.

CMAES has one critical downside: it has a limited exploration capacity
due to the highly localized nature of the algorithm.
To mitigate this, similarly to Ref.~\cite{Romao:2024gjx},
we implemented a novelty reward into the loss function.
This is done by adding the density of found points as a penalty to the loss function,
thus allowing it to be minimal when the point is good and away from other good points.
To this end, we used the Histogram Based Outlier System (HBOS),
which computes a density penalty between 0 and 1,
such that a new point has 0 and a point similar to previous ones approach
the maximum 1, basing itself on the abundance
values for chosen parameters.

To ensure proper minimization with these penalties, the loss function is shifted accordingly,
\begin{equation}
    \Tilde{L}(\theta) =
    \left\{
    \begin{array}{ll}
        1 + L(\theta)   & \text{if } L(\theta) > 0 \\
        \text{ } \\
        0               & \text{if } L(\theta) = 0
    \end{array}
    \right..
\end{equation}
The penalties are then added to obtain the final version of the loss function,
\begin{equation}
    L_T(\theta) = \Tilde{L}(\theta) + \frac{1}{2} \left( \frac{1}{N^{\mathcal{P}}_p} \sum_{i=1}^{N^{\mathcal{P}}_p} p^{\mathcal{P}}_i(\theta^i) + \frac{1}{N^{\mathcal{O}}_p}\sum_{i=1}^{N^{\mathcal{O}}_p}p^{\mathcal{O}}_i [\mathcal{O}^{i}(\theta)]\right),
\end{equation}
where $p^{\mathcal{P}}_i(\theta^i)$ is the density penalty in parameter
space $\mathcal{P}$, normalized by the amount of parameter penalties
considered, $N^{\mathcal{P}}_p$, and $p^{\mathcal{O}}_i [\mathcal{O}^{i}(\theta)]$
is the density penalty of the observable space $\mathcal{O}$,
also properly normalized by $N^{\mathcal{O}}_p$.
We highlight the fact that penalties do not need to apply to all parameters $\theta$
and/or observables $\mathcal{O}(\theta)$.
A subset of interest can be chosen to perform \textit{focused runs} with
density penalty on  specific parameters and/or observables.

With the default setup, each optimization run is independent,
as CMA-ES is  initialized with new values for the  mean and parameters in the covariant
matrix and trained solely on points from that run.
We may however choose valid points from previous runs as seeds to start new runs with CMA-ES initialized
already in that region.

\begin{figure}[htb]
	\centering
	\includegraphics[height=5cm]{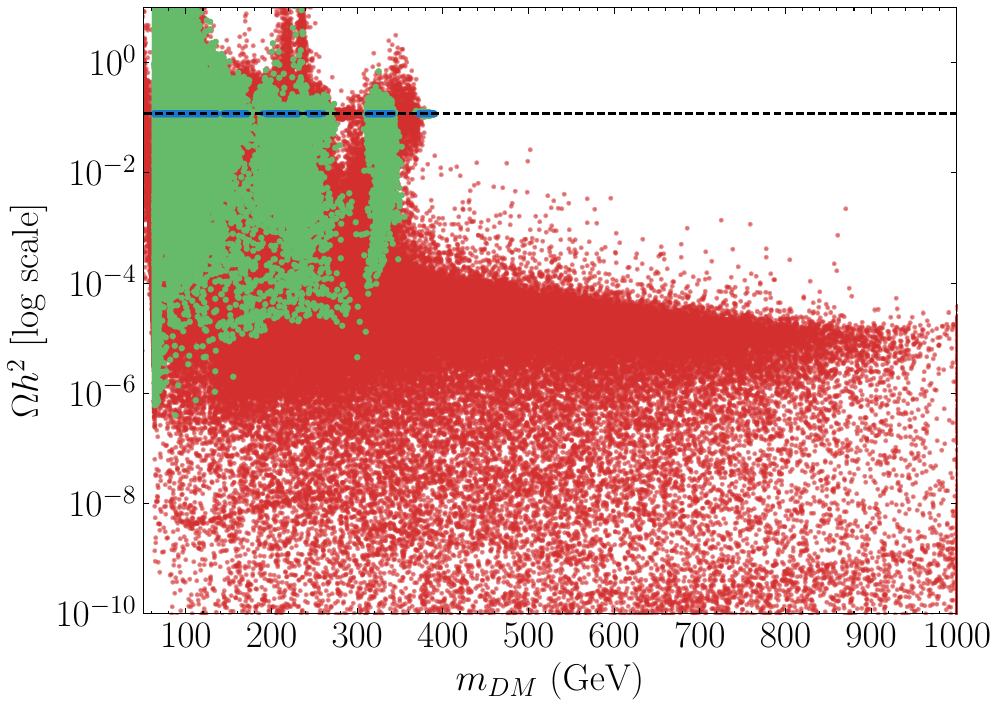}
	\includegraphics[height=5cm]{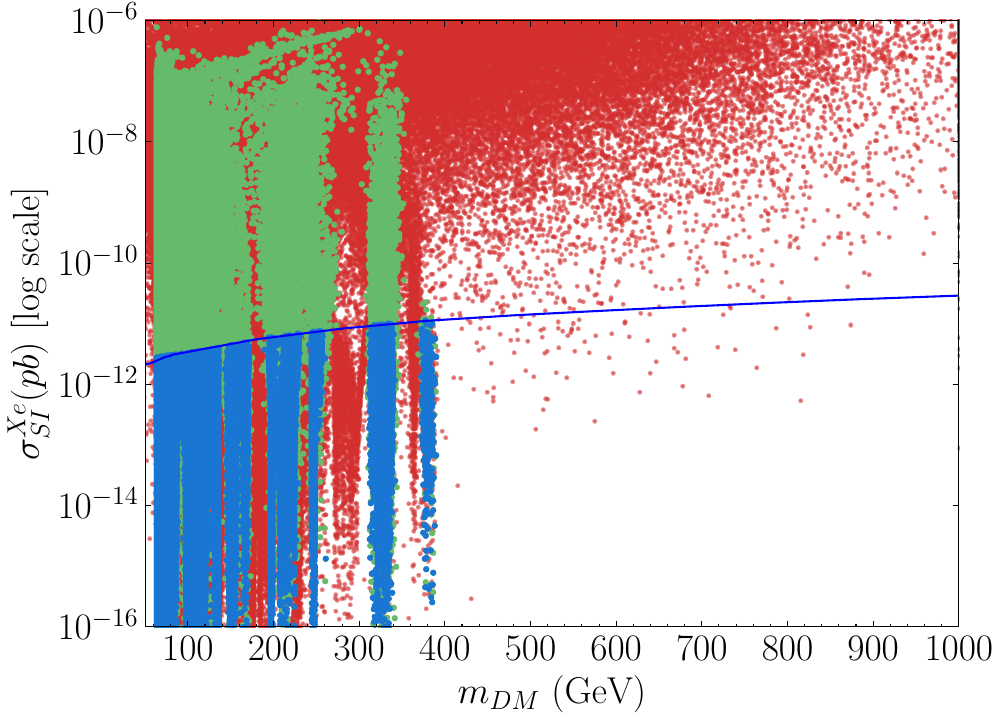}
\caption{\label{omega_nofocus} Results from a run using the optimization
algorithm CMA-ES, without the novelty reward or seeded runs methods
described in the text. In the Left panel we show the rapid convergence towards
the imposed interval on the total relic density, $\Omega_T h^2 \in [0.1164,0.1236]$,
and in the Right panel the obtained value for the direct detection quantity
$\sigma_{\text{SI}}^{\text{Xe}}$ given by eq.~\eqref{sigmascat}, with the most recent
constraint from LZ~\cite{LZ:2024zvo}, shown as the blue line.
The points in red fail one of the constraints not related to DM and the points in green meet all constraints except DM. The points with a blue color are points in green that also satisfy the correct relic density and direct detection scattering bounds.}
\end{figure}

We implemented the optimization with CMA-ES and performed a first simulation,
including a $C(\mathcal{O})$ for the relic density,
without novelty reward and not considering bounds from DM direct/indirect searches.
On $\sim 1500$ CPU hours, we obtained $\sim 660000$ points satisfying
$\Omega_T h^2 \in [0.1164,0.1236]$,
shown in Fig.~\ref{omega_nofocus},
for the same plane as in Figs.~\ref{omega_random} and \ref{direct_k10}. The points in \textit{red} fail one of the constraints in Sections~\ref{BFB}-~\ref{sec:exp}, identified by least one $C(\mathcal{O})$ in the loss function not zero.
The points in \textit{green} are found to satisfy all constraints apart from DM bounds, satisfying BFB, unitarity, global minimum, flavour bounds, coupling modifiers and  \texttt{HiggsTools-1.1.3}.
In \textit{blue} are the points originally in green found
to also meet the Dark Matter constraints.
The iterative process quickly moves towards the regions with considerable
relic density and we are able to obtain points also satisfying direct detection bounds.
However, the lack of novelty reward results in a large concentration of points in
the same mass region for the DM candidate, around $65$-$300\textrm{GeV}$.
In the next section we follow the method with novelty reward,
that considers the relevant DM bounds with appropriate $C(\mathcal{O})$
for each experimental exclusion.

\section{\label{sec:results}Final Machine Learning Results}

Following the initial set of simulations showing the ability
of the evolutionary algorithm in obtaining parameter points with the
correct relic density, but lacking diversity in the Dark Matter mass plane,
we set our goal on fully exploring the allowed parameter space adding the novelty reward technique.

When we include indirect constraints, with exclusion lines shown in
Fig.~\ref{indirect_lines},
we observe an increase in the time per generation
of a factor of 4 and an increased convergence time.
The strategy followed was then to only use
\texttt{micrOMEGAs-6.2.3}~\cite{Alguero:2023zol} to numerically calculate
the relic density and direct detection variables.
For a set of simulations, we took parameter points that met the relic
density from Planck and nucleon scattering constraints to check the annihilation bounds.
As a next step,
we passed again the valid (relic and direct detection) points through
\texttt{micrOMEGAs-6.2.3}, probing for consistency with indirect detection bounds.
We found the large majority of points to already satisfy the current
indirect detection constraints.
The final results, also satisfying all other constraints
described in Sections~\ref{BFB}-~\ref{sec:exp}, have the Dark Matter relic density within $3\sigma$ of the limits obtained
by the Planck experiment~\cite{Planck:2018vyg}, satisfy the most recent direct
detection bounds from LZ~\cite{LZ:2024zvo}, as well as the indirect detection constraints
from Fig.~\ref{indirect_lines}.
All other constraints described in Sections~\ref{BFB}-~\ref{sec:exp} are also 
considered.  

We start with some run,
which typically will yield one of the line
structures visible in Fig.~\ref{omega_nofocus}.
Next, we take a set of valid points,
and we start seeded runs based on those,
specifically targeting a novel mass region.
For example, we looked specially into regions where the DM candidate lies below the
mass of $m_{h1} = 125\,\text{GeV}$ and also regions above $500~\text{GeV}$.
We believe that the gaps in the mass plane can be filled completely in
the mass region shown. The allowed mass region for the DM candidate spans
the entire mass range from half the Higgs mass to the high GeV scale.

No valid points were found in the mass region below $m_{h_1}/2$,
as the latest LHC bound on the invisible branching ratio
of the Higgs, BR($h\rightarrow \text{inv}$) $\leq$ 0.107 at $95\%$
C.L.~\cite{ATLAS:2023tkt}, comes into effect.

\begin{figure}[htb]
	\centering
	\includegraphics[height=5cm]{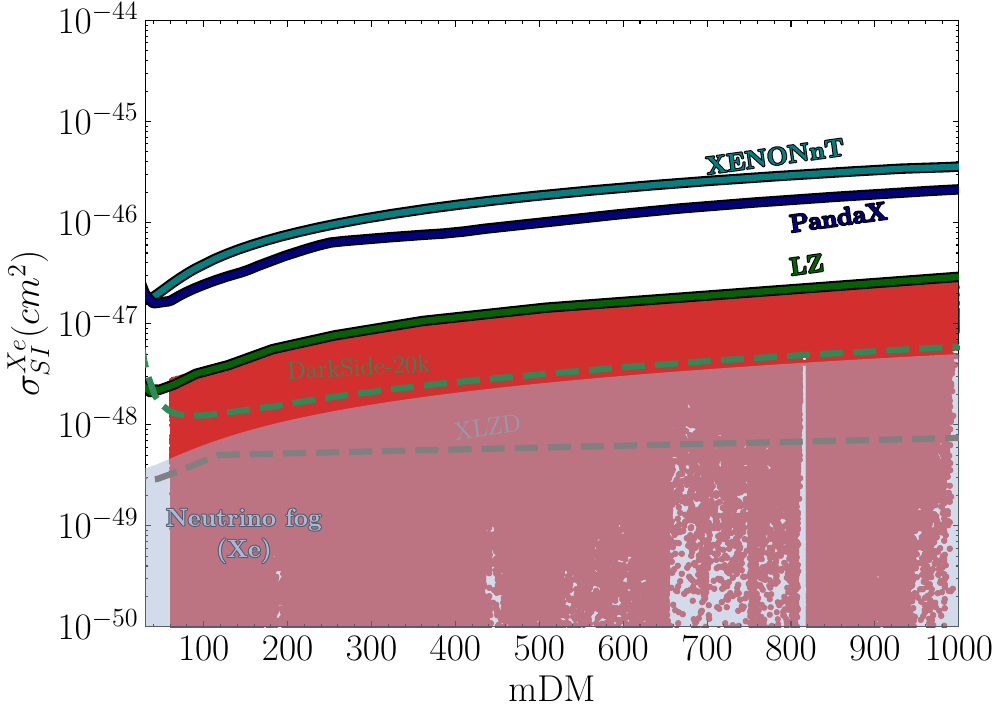}
\caption{\label{massfocus_direct} The red region correponds to valid points generated with our model.
The solid lines shown correspond to the most recent exclusion bounds
from experiments, with XENONnT~\cite{XENON:2025vwd}, PandaX-4T~\cite{PandaX:2024qfu}
and LZ~\cite{LZ:2024zvo}.
The dashed lines show the projections for DarkSide-20k~\cite{DarkSide-20k:2017zyg}
and XLZD~\cite{XLZD:2024nsu}.
The neutrino floor is shown in grey, 
as defined in Ref.~\cite{OHare:2021utq}.}
\end{figure}

In Fig.~\ref{massfocus_direct},
we show the constraints from direct
detection\footnote{The data files and general layout for the plots was
obtained from the public repository by Ciaran O'Hare avaliable
in \url{https://github.com/cajohare/DirectDetectionPlots}.} for the mass region studied.
The solid lines shown correspond to the most recent exclusion bounds
from experiments, with XENONnT~\cite{XENON:2025vwd}, PandaX-4T~\cite{PandaX:2024qfu}
and LZ~\cite{LZ:2024zvo}.
The dashed lines show the projections for DarkSide-20k~\cite{DarkSide-20k:2017zyg}
and XLZD~\cite{XLZD:2024nsu}.
The neutrino floor is shown in grey, 
as defined in Ref.~\cite{OHare:2021utq}.

As one can see from Fig.~\ref{massfocus_direct},
there are many valid points obeying the current LZ bounds
(in addition to matching the correct relic density).
Some of those points could be excluded (or confirmed) by future experiments, such as 
DarkSide-20k or XLZD.
However,
the fact that many red (valid model) points lie within the grey (neutrino fog) region,
implies that this model will be hard to exclude with future direct detection
experiments~\footnote{There is also the possibility of improving theoretical
and experimental tools in order to break the degeneracy between neutrino
backgrounds and potential WIMP signals
\cite{Davis:2014ama, Gelmini:2018ogy, Arcadi:2025ifl, Vahsen:2020pzb}.}.

\section{\label{sec:conclusion}Conclusion}

We have studied in detail a model with a type II 2HDM supplemented by two inert scalars which
provide a viable dark matter particle.
We developed here for the first time a full study of:
the possible vaccua;
the (sufficient) bounded from below requirements;
the conditions under which our chosen vacuum
- $\mathcal{N}$ in eq.~\eqref{vev_nosso} - is the global minimum;
and the perturbative unitarity constraints.

Points in parameter space where searched for,
which obeyed these conditions and, in addition,
satisfy current experimental bounds arising from the oblique radiative parameters,
current LHC bounds on the $125\textrm{GeV}$ Higgs as well as searches for additional scalar particles,
and also direct, indirect, and relic abundance constraints.
These constraints have been implemented using \texttt{HiggsTools-1.1.3}~\cite{Bahl:2022igd} and
\texttt{micrOMEGAs-6.2.3}~\cite{Alguero:2023zol}.

Three different scanning strategies have been utilized:
i)\, a random scan without any prior assumptions on the parameter space;
ii)\, a scan close to the alignment limit of the 2HDM,
defined as $\alpha_1=-\beta$ for the choices in  eqs.~(\ref{obeta}) and (\ref{field_eq});
iii)\, employing the Artificial Intelligence black box optimization approach.
We have found that the random scan and the scan close to the alignment limit of the 2HDM produce
similar results, although (of course) the latter is more efficient than the first,
by a factor close to three.
Neither of these produce points with the correct relic density in a reasonable time frame.
Thus, we turned to a machine learning method.
Using unseeded runs,
we quickly find solutions with reasonable relic density,
but yielding limited exploration of the various mass regions.
In contrast, using a novelty reward strategy and seeded runs,
we are able to find valid points in a larger whole mass region:
$125\textrm{GeV}/2 < m_{DM} \lesssim 1000\textrm{GeV}$.

We have found that current/projected direct detection experiments do/will not
exclude this model, since there are many valid points with the neutrino fog region.
We have also studied current indirect detection constraints,
and found that they do not significantly affect those points already
allowed by direct detection bounds and relic density.

All models addressing the DM problem with scalar DM candidates must contend
with a plethora of experimental results,
from collider and astrophysical searches.
When models have many parameters, such searches are very demanding computationally.
One can easily seem to rule out a valid model or be tempted to take
as a generic model feature, properties which were merely the result of
searches performed in a region where valid points are easy to produce. 
This work illustrates the importance of using Machine Learning coupled with
novelty detection techniques in order to efficiently explore the validity and
generic features of DM models where extensive collider data must also be
contended with.
Harnessing the power of Machine Learning and novelty detection,
opens the door to deeper, more reliable insights into dark matter
models—paving the way for discovery in even the most complex parameter spaces.

\section{Acknowledgments}

This work is supported in part by the Portuguese
Fundação para a Ciência e Tecnologia (FCT) through the PRR (Recovery and Resilience
Plan), within the scope of the investment "RE-C06-i06 - Science Plus Capacity Building", measure "RE-C06-i06.m02 - Reinforcement of financing for International Partnerships in Science,
Technology and Innovation of the PRR", under the project with reference 2024.01362.CERN.
The work of the authors is
also supported by FCT under Contracts UIDB/00777/2020, and UIDP/00777/2020.
The FCT projects are partially funded through
POCTI (FEDER), COMPETE, QREN, and the EU.
The work of R. Boto is also supported by FCT with the PhD
grant PRT/BD/152268/2021.

\appendix

\section{Explicit expressions for depth of minimum}

For the  desired $\mathcal{N}$-type vacuum to be stable, it must be the global minimum of the potential. To verify that, we used equation (\ref{min-comparison}) to compare this vacuum with the other minima expressed in section \ref{Formalism and Vacua} and arrived at the following conclusions.

\subsection{\texorpdfstring{$\mathcal{N}$-type vacuum}{N-type vacuum}}
The expressions for comparing the depth of the $\mathcal{N}$-type potential against $\mathcal{CB}$ and $\mathcal{CP}$-type vacua are:
\begin{align}
\mathcal{V}_{\mathcal{CB}} - \mathcal{V}_{\mathcal{N}} = &\left(\frac{m_{H^\pm}^2}{4v^2}\right)_\mathcal{N} 
\big[(v_2 c_1 - v_1 c_3)^2 + v_1^2 c_2^2 \big] , \\
\mathcal{V}_{\mathcal{CB}_s} - \mathcal{V}_{\mathcal{N}} =& \left(\frac{m_{H^\pm}^2}{4v^2}\right)_\mathcal{N} 
\big[(v_2 c'_1 - v_1 c'_3)^2 + v_1^2 {c'_2}^2\big] + \frac{1}{4} {c'_S}^2 \big(m_s^2\big)_{\mathcal{N}} , \\
\mathcal{V}_{\mathcal{CB}_p} - \mathcal{V}_{\mathcal{N}} =& \left(\frac{m_{H^\pm}^2}{4v^2}\right)_\mathcal{N} 
\big[(v_2 c''_1 - v_1 c''_3)^2 + v_1^2 {c''_2}^2\big] +\frac{1}{4} {c''_P}^2 \big(m_p^2\big)_{\mathcal{N}} , \\
\mathcal{V}_{\mathcal{CB}_{sp}} - \mathcal{V}_{\mathcal{N}} =& \left(\frac{m_{H^\pm}^2}{4v^2}\right)_\mathcal{N} 
\big[(v_2 c'''_1 - v_1 c'''_3)^2 + v_1^2 {c'''_2}^2\big] +\frac{1}{4} {c'''_S}^2 \big(m_s^2\big)_{\mathcal{N}} +\frac{1}{4} {c'''_P}^2 \big(m_p^2\big)_{\mathcal{N}}  \notag\\
&+\frac{1}{2} c'''_S c'''_P \big(m_{sp}^2\big)_{\mathcal{N}}, \\
\mathcal{V}_{\mathcal{CP}} - \mathcal{V}_{\mathcal{N}} =& \left(\frac{m_{A}^2}{4v^2}\right)_\mathcal{N} 
\big[(v_2 \bar{v}_1 - v_1 \bar{v}_2)^2 + v_1^2 \bar{v}_3^2 \big] , \\
\mathcal{V}_{\mathcal{CP}_s} - \mathcal{V}_{\mathcal{N}} =& \left(\frac{m_{A}^2}{4v^2}\right)_\mathcal{N} 
\big[(v_2 \bar{v}'_1 - v_1 \bar{v}'_2)^2 + v_1^2 {\left.\bar{v}'_3\right.}^2 \big] + \frac{1}{4} {\left.\bar{v}'_S\right.}^2 \big(m_s^2\big)_{\mathcal{N}} , \\
\mathcal{V}_{\mathcal{CP}_p} - \mathcal{V}_{\mathcal{N}} =& \left(\frac{m_{A}^2}{4v^2}\right)_\mathcal{N} 
\big[\left(v_2 \bar{v}''_1 - v_1 \bar{v}''_2\right)^2 + v_1^2 {\left.\bar{v}''_3\right.}^2 \big] +\frac{1}{4} {\left.\bar{v}''_P\right.}^2 \big(m_p^2\big)_{\mathcal{N}} , \\
\mathcal{V}_{\mathcal{CP}_{sp}} - \mathcal{V}_{\mathcal{N}} =& \left(\frac{m_{A}^2}{4v^2}\right)_\mathcal{N} 
\big[(v_2 \bar{v}'''_1 - v_1 \bar{v}'''_2)^2 + v_1^2 {\left.\bar{v}'''_3\right.}^2 \big] +\frac{1}{4}{\left.\bar{v}'''_S\right.}^2 \big(m_s^2\big)_{\mathcal{N}} +\frac{1}{4} {\left.\bar{v}'''_P\right.}^2 \big(m_p^2\big)_{\mathcal{N}} \notag \\
&+ \frac{1}{2} \bar{v}'''_S \bar{v}'''_P \big(m_{sp}^2\big)_{\mathcal{N}}.
\end{align}

From these equations, one can infer that \textit{if the potential has a
minimum of type $\mathcal{N}$, any stationary point of type
$\mathcal{CB}, \mathcal{CB}_s, \mathcal{CB}_p, \mathcal{CP}, \mathcal{CP}_s$
or $\mathcal{CP}_p$, if it exists, lies above $\mathcal{N}$.
This is not necessarily the case for stationary points of type
$\mathcal{CB}_{sp}$ or $\mathcal{CP}_{sp}$.}

\subsection{\texorpdfstring{$\mathcal{N}_s$-type vacuum}{Ns-type vacuum}}

The expressions for comparing the depth of the $\mathcal{N}_s$-type potential against $\mathcal{CB}$ and $\mathcal{CP}$-type vacua are:

\begin{align}
\mathcal{V}_{\mathcal{CB}} - \mathcal{V}_{\mathcal{N}_s} = &\left(\frac{m_{H^\pm}^2}{4{v'}^2}\right)_{\mathcal{N}_s}
\big[(v'_2 c_1 - v'_1 c_3)^2 + {v'_1}^2 c_2^2 \big] - \frac{1}{4} {v'_S}^2 \big(m_s^2\big)_{\mathcal{CB}} , \\
\mathcal{V}_{\mathcal{CB}_s} - \mathcal{V}_{\mathcal{N}_s} =& \left(\frac{m_{H^\pm}^2}{4{v'}^2}\right)_{\mathcal{N}_s} 
\big[(v'_2 c'_1 - v'_1 c'_3)^2 + {v'_1}^2 {c'_2}^2\big], \\
\mathcal{V}_{\mathcal{CB}_p} - \mathcal{V}_{\mathcal{N}_s} =& \left(\frac{m_{H^\pm}^2}{4{v'}^2}\right)_{\mathcal{N}_s} 
\big[(v'_2 c''_1 - v'_1 c''_3)^2 + {v'_1}^2 {c''_2}^2\big] +\frac{1}{4} {c''_P}^2 \big(m_p^2\big)_{\mathcal{N}_s} - \frac{1}{4} {v'_S}^2 \big(m_s^2\big)_{\mathcal{CB}_p}, \\
\mathcal{V}_{\mathcal{CB}_{sp}} - \mathcal{V}_{\mathcal{N}_s} =& \left(\frac{m_{H^\pm}^2}{4{v'}^2}\right)_{\mathcal{N}_s} 
\big[(v'_2 c'''_1 - v'_1 c'''_3)^2 + {v'_1}^2 {c'''_2}^2\big] +\frac{1}{4} {c'''_P}^2 \big(m_p^2\big)_{\mathcal{N}_s} +\frac{1}{2} c'''_S c'''_P \big(m_{sp}^2\big)_{\mathcal{N}_s}  \notag\\
& - \frac{1}{4} {v'_S}^2 \big(m_s^2\big)_{\mathcal{CB}_{sp}}, \\
\mathcal{V}_{\mathcal{CP}} - \mathcal{V}_{\mathcal{N}_s} =& \left(\frac{m_{A}^2}{4{v'}^2}\right)_{\mathcal{N}_s} 
\big[(v'_2 \bar{v}_1 - v'_1 \bar{v}_2)^2 + {v'_1}^2 \bar{v}_3^2 \big] - \frac{1}{4} {v'_S}^2 \big(m_s^2\big)_{\mathcal{CP}}, \\
\mathcal{V}_{\mathcal{CP}_s} - \mathcal{V}_{\mathcal{N}_s} =& \left(\frac{m_{A}^2}{4{v'}^2}\right)_{\mathcal{N}_s} 
\big[(v'_2 \bar{v}'_1 - v'_1 \bar{v}'_2)^2 + {v'_1}^2 {\left.\bar{v}'_3\right.}^2 \big] , \\
\mathcal{V}_{\mathcal{CP}_p} - \mathcal{V}_{\mathcal{N}_s} =& \left(\frac{m_{A}^2}{4{v'}^2}\right)_{\mathcal{N}_s}
\big[(v'_2 \bar{v}''_1 - v'_1 \bar{v}''_2)^2 + {v'_1}^2 {\left.\bar{v}''_3\right.}^2 \big] +\frac{1}{4} {\left.\bar{v}''_P\right.}^2 \big(m_p^2\big)_{\mathcal{N}_s} - \frac{1}{4} {v'_S}^2 \big(m_s^2\big)_{\mathcal{CP}_p} , \\
\mathcal{V}_{\mathcal{CP}_{sp}} - \mathcal{V}_{\mathcal{N}_s} =& \left(\frac{m_{A}^2}{4{v'}^2}\right)_{\mathcal{N}_s} 
\big[(v'_2 \bar{v}'''_1 - v'_1 \bar{v}'''_2)^2 + {v'_1}^2 {\left.\bar{v}'''_3\right.}^2 \big] +\frac{1}{4} {\left.\bar{v}'''_P\right.}^2 \big(m_p^2\big)_{\mathcal{N}_s} +\frac{1}{2} \bar{v}'''_S \bar{v}'''_P \big(m_{sp}^2\big)_{\mathcal{N}_s}  \notag\\
& - \frac{1}{4} {v'_S}^2 \big(m_s^2\big)_{\mathcal{CP}_{sp}}.
\end{align}

From these expressions, we conclude that \textit{if the potential
has a minimum of type $\mathcal{N}_s$, any stationary point of type
$\mathcal{CB}_s$ or $\mathcal{CP}_s$, if it exists, lies above $\mathcal{N}_s$.
This is not necessarily the case for the other stationary points of
type $\mathcal{CB}$ and $\mathcal{CP}$}.

\subsection{\texorpdfstring{$\mathcal{N}_p$-type vacuum}{Np-type vacuum}}

The expressions for comparing the depth of the $\mathcal{N}_p$-type
potential against $\mathcal{CB}$ and $\mathcal{CP}$-type vacua are:
\begin{align}
\mathcal{V}_{\mathcal{CB}} - \mathcal{V}_{\mathcal{N}_p} = &\left(\frac{m_{H^\pm}^2}{4{v''}^2}\right)_{\mathcal{N}_p} 
\big[(v''_2 c_1 - v''_1 c_3)^2 + {v''_1}^2 c_2^2 \big] - \frac{1}{4} {v''_P}^2 \big(m_p^2\big)_{\mathcal{CB}} , \\
\mathcal{V}_{\mathcal{CB}_s} - \mathcal{V}_{\mathcal{N}_p} =& \left(\frac{m_{H^\pm}^2}{4{v''}^2}\right)_{\mathcal{N}_p} 
\big[(v''_2 c'_1 - v''_1 c'_3)^2 + {v''_1}^2 {c'_2}^2\big] +\frac{1}{4} {c'_S}^2 \big(m_S^2\big)_{\mathcal{N}_p} - \frac{1}{4} {v''_P}^2 \big(m_p^2\big)_{\mathcal{CB}_s}, \\
\mathcal{V}_{\mathcal{CB}_p} - \mathcal{V}_{\mathcal{N}_p} =& \left(\frac{m_{H^\pm}^2}{4{v''}^2}\right)_{\mathcal{N}_p} 
\big[(v''_2 c''_1 - v''_1 c''_3)^2 + {v''_1}^2 {c''_2}^2\big], \\
\mathcal{V}_{\mathcal{CB}_{sp}} - \mathcal{V}_{\mathcal{N}_p} =& \left(\frac{m_{H^\pm}^2}{4{v''}^2}\right)_{\mathcal{N}_p} 
\big[(v''_2 c'''_1 - v''_1 c'''_3)^2 + {v''_1}^2 {c'''_2}^2\big] +\frac{1}{4} {c'''_S}^2 \big(m_s^2\big)_{\mathcal{N}_p} +\frac{1}{2} c'''_S c'''_P \big(m_{sp}^2\big)_{\mathcal{N}_p}  \notag\\
& - \frac{1}{4} {v''_P}^2 \big(m_p^2\big)_{\mathcal{CB}_{sp}}, \\
\mathcal{V}_{\mathcal{CP}} - \mathcal{V}_{\mathcal{N}_p} =& \left(\frac{m_{A}^2}{4{v''}^2}\right)_{\mathcal{N}_p} 
\big[(v''_2 \bar{v}_1 - v''_1 \bar{v}_2)^2 + {v''_1}^2 \bar{v}_3^2 \big] - \frac{1}{4} {v''_P}^2 \big(m_p^2\big)_{\mathcal{CP}}, \\
\mathcal{V}_{\mathcal{CP}_s} - \mathcal{V}_{\mathcal{N}_p} =& \left(\frac{m_{A}^2}{4{v''}^2}\right)_{\mathcal{N}_p} 
\big[(v''_2 \bar{v}'_1 - v''_1 \bar{v}'_2)^2 + {v''_1}^2 {\left.\bar{v}'_3\right.}^2 \big] +\frac{1}{4} {\left.\bar{v}'_S\right.}^2 \big(m_s^2\big)_{\mathcal{N}_p} - \frac{1}{4} {v''_P}^2 \big(m_p^2\big)_{\mathcal{CP}_s}, \\
\mathcal{V}_{\mathcal{CP}_p} - \mathcal{V}_{\mathcal{N}_p} =& \left(\frac{m_{A}^2}{4{v''}^2}\right)_{\mathcal{N}_p}
\big[(v''_2 \bar{v}''_1 - v''_1 \bar{v}''_2)^2 + {v''_1}^2 {\left.\bar{v}''_3\right.}^2 \big]  , \\
\mathcal{V}_{\mathcal{CP}_{sp}} - \mathcal{V}_{\mathcal{N}_p} =& \left(\frac{m_{A}^2}{4{v''}^2}\right)_{\mathcal{N}_p} 
\big[(v''_2 \bar{v}'''_1 - v''_1 \bar{v}'''_2)^2 + {v''_1}^2 {\left.\bar{v}'''_3\right.}^2 \big] +\frac{1}{4} {\left.\bar{v}'''_S\right.}^2 \big(m_s^2\big)_{\mathcal{N}_p} +\frac{1}{2} \bar{v}'''_S \bar{v}'''_P \big(m_{sp}^2\big)_{\mathcal{N}_p}  \notag\\
& - \frac{1}{4} {v''_P}^2 \big(m_p^2\big)_{\mathcal{CP}_{sp}}.
\end{align}
Analogously to the previous case, \textit{if the potential has a minimum of
type $\mathcal{N}_p$, any stationary point of type $\mathcal{CB}_p$
or $\mathcal{CP}_p$, if it exists, lies above $\mathcal{N}_p$.
This is not necessarily the case for the other stationary points of
type $\mathcal{CB}$ and $\mathcal{CP}$}.

\subsection{\texorpdfstring{$\mathcal{N}_{sp}$-type vacuum}{Nsp-type vacuum}}

The expressions for comparing the depth of the $\mathcal{N}_{sp}$-type potential against $\mathcal{CB}$ and $\mathcal{CP}$-type vacua are:
\begin{align}
\mathcal{V}_{\mathcal{CB}} - \mathcal{V}_{\mathcal{N}_{sp}} = &\left(\frac{m_{H^\pm}^2}{4{v'''}^2}\right)_{\mathcal{N}_{sp}} 
\big[(v'''_2 c_1 - v'''_1 c_3)^2 + {v'''_1}^2 c_2^2 \big] - \frac{1}{4} {v'''_S}^2 \big(m_s^2 \big)_{\mathcal{CB}} - \frac{1}{4} {v'''_P}^2 \big(m_p^2 \big)_{\mathcal{CB}} \notag \\
& - \frac{1}{2} v'''_S v'''_P \big(m_{sp}^2 \big)_{\mathcal{CB}}, \\
\mathcal{V}_{\mathcal{CB}_s} - \mathcal{V}_{\mathcal{N}_{sp}} = &\left(\frac{m_{H^\pm}^2}{4{v'''}^2}\right)_{\mathcal{N}_{sp}} 
\big[(v'''_2 c'_1 - v'''_1 c'_3)^2 + {v'''_1}^2 {c'_2}^2 \big] + \frac{1}{4} {c'_S}^2 \big(m_s^2 \big)_{\mathcal{N}_{sp}} - \frac{1}{4} {v'''_P}^2 \big(m_p^2 \big)_{\mathcal{CB}_s} \notag \\
& - \frac{1}{2} v'''_S v'''_P \big(m_{sp}^2 \big)_{\mathcal{CB}_s}, \\
\mathcal{V}_{\mathcal{CB}_p} - \mathcal{V}_{\mathcal{N}_{sp}} = &\left(\frac{m_{H^\pm}^2}{4{v'''}^2}\right)_{\mathcal{N}_{sp}} 
\big[(v'''_2 c''_1 - v'''_1 c''_3)^2 + {v'''_1}^2 {c''_2}^2 \big] + \frac{1}{4} {c''_P}^2 \big(m_p^2 \big)_{\mathcal{N}_{sp}} - \frac{1}{4} {v'''_S}^2 \big(m_s^2 \big)_{\mathcal{CB}_p}  \notag \\
& - \frac{1}{2} v'''_S v'''_P \big(m_{sp}^2 \big)_{\mathcal{CB}_p}, \\
\mathcal{V}_{\mathcal{CB}_{sp}} - \mathcal{V}_{\mathcal{N}_{sp}} = &\left(\frac{m_{H^\pm}^2}{4{v'''}^2}\right)_{\mathcal{N}_{sp}} 
\big[(v'''_2 c'''_1 - v'''_1 c'''_3)^2 + {v'''_1}^2 {c'''_2}^2 \big] + \frac{1}{4} {c'''_S}^2 \big(m_s^2 \big)_{\mathcal{N}_{sp}} + \frac{1}{4} {c'''_P}^2 \big(m_p^2 \big)_{\mathcal{N}_{sp}} \notag \\
& + \frac{1}{2} c'''_S c'''_P \big(m_{sp}^2 \big)_{\mathcal{N}_{sp}} - \frac{1}{4} {v'''_S}^2 \big(m_s^2 \big)_{\mathcal{CB}_{sp}} - \frac{1}{4} {v'''_P}^2 \big(m_p^2 \big)_{\mathcal{CB}_{sp}} - \frac{1}{2} v'''_S v'''_P \big(m_{sp}^2 \big)_{\mathcal{CB}_p} , \\
\mathcal{V}_{\mathcal{CP}} - \mathcal{V}_{\mathcal{N}_{sp}} =& \left(\frac{m_{A}^2}{4{v'''}^2}\right)_{\mathcal{N}_{sp}} 
\big[(v'''_2 \bar{v}_1 - v'''_1 \bar{v}_2)^2 + {v'''_1}^2 \bar{v}_3^2 \big] - \frac{1}{4} {v'''_S}^2 \big(m_s^2 \big)_{\mathcal{CP}} - \frac{1}{4} {v'''_P}^2 \big(m_p^2 \big)_{\mathcal{CP}} \notag \\
& - \frac{1}{2} v'''_S v'''_P \big(m_{sp}^2 \big)_{\mathcal{CP}}, \\
\mathcal{V}_{\mathcal{CP}_s} - \mathcal{V}_{\mathcal{N}_{sp}} =& \left(\frac{m_{A}^2}{4{v'''}^2}\right)_{\mathcal{N}_{sp}} 
\big[(v'''_2 \bar{v}'_1 - v'''_1 \bar{v}'_2)^2 + {v'''_1}^2 {\left.\bar{v}'_3\right.}^2 \big] + \frac{1}{4} {\left.\bar{v}'_S\right.}^2 \big(m_s^2 \big)_{\mathcal{N}_{sp}} - \frac{1}{4} {v'''_P}^2 \big(m_p^2 \big)_{\mathcal{CP}_s} \notag \\
& - \frac{1}{2} v'''_S v'''_P \big(m_{sp}^2 \big)_{\mathcal{CP}_s}, \\
\mathcal{V}_{\mathcal{CP}_p} - \mathcal{V}_{\mathcal{N}_{sp}} =& \left(\frac{m_{A}^2}{4{v'''}^2}\right)_{\mathcal{N}_{sp}} 
\big[(v'''_2 \bar{v}''_1 - v'''_1 \bar{v}''_2)^2 + {v'''_1}^2 {\left.\bar{v}''_3\right.}^2 \big] + \frac{1}{4} {\left.\bar{v}''_P\right.}^2 \big(m_p^2 \big)_{\mathcal{N}_{sp}} - \frac{1}{4} {v'''_S}^2 \big(m_s^2 \big)_{\mathcal{CP}_p}  \notag \\
& - \frac{1}{2} v'''_S v'''_P \big(m_{sp}^2 \big)_{\mathcal{CP}_p} , \\
\mathcal{V}_{\mathcal{CP}_{sp}} - \mathcal{V}_{\mathcal{N}_{sp}} =& \left(\frac{m_{A}^2}{4{v'''}^2}\right)_{\mathcal{N}_{sp}} 
\big[(v'''_2 \bar{v}'''_1 - v'''_1 \bar{v}'''_2)^2 + {v'''_1}^2 {\left.\bar{v}'''_3\right.}^2 \big]  + \frac{1}{4} {\left.\bar{v}'''_S\right.}^2 \big(m_s^2 \big)_{\mathcal{N}_{sp}} + \frac{1}{4} {\left.\bar{v}'''_P\right.}^2 \big(m_p^2 \big)_{\mathcal{N}_{sp}} \notag \\
& + \frac{1}{2} \bar{v}'''_S \bar{v}'''_P \big(m_{sp}^2 \big)_{\mathcal{N}_{sp}} - \frac{1}{4} {v'''_S}^2 \big(m_s^2 \big)_{\mathcal{CP}_{sp}} - \frac{1}{4} {v'''_P}^2 \big(m_p^2 \big)_{\mathcal{CP}_{sp}} - \frac{1}{2} v'''_S v'''_P \big(m_{sp}^2 \big)_{\mathcal{CP}_p} .
\end{align}
\textit{There is no guarantee that the $\mathcal{N}_{sp}$ vacuum is deeper than any $\mathcal{CB}$ or $\mathcal{CP}$-type vacuum}.

\subsection{Coexisting Neutral minima}

Another contingency that we must take is to verify the stability between $\mathcal{N}, \mathcal{N}_s, \mathcal{N}_p$ and $\mathcal{N}_{sp}$. The equations relating their potential depth are the following.

\begin{align}
    \mathcal{V}_{\mathcal{N}_s} - \mathcal{V}_{\mathcal{N}} = &\left[\left(\frac{m_{H^\pm}^2}{4v^2}\right)_{\mathcal{N}} - \left(\frac{m_{H^\pm}^2}{4{v'}^2}\right)_{\mathcal{N}_s}\right](v_2'v_1 - v_1'v_2)^2 +\frac{1}{4} {v'_S}^2 \big(m_s^2\big)_{\mathcal{N}}, \\
    \mathcal{V}_{\mathcal{N}_p} - \mathcal{V}_{\mathcal{N}} =& \left[\left(\frac{m_{H^\pm}^2}{4v^2}\right)_{\mathcal{N}} - \left(\frac{m_{H^\pm}^2}{4{v''}^2}\right)_{\mathcal{N}_p}\right](v_2''v_1 - v_1''v_2)^2 +\frac{1}{4} {v''_P}^2 \big(m_p^2\big)_{\mathcal{N}}, \\
    \mathcal{V}_{\mathcal{N}_{sp}} - \mathcal{V}_{\mathcal{N}} =& \left[\left(\frac{m_{H^\pm}^2}{4v^2}\right)_{\mathcal{N}} - \left(\frac{m_{H^\pm}^2}{4{v'''}^2}\right)_{\mathcal{N}_{sp}}\right](v_2'''v_1 - v_1'''v_2)^2 +\frac{1}{4} {v'''_S}^2 \big(m_s^2\big)_{\mathcal{N}} +\frac{1}{4} {v'''_P}^2 \big(m_p^2\big)_{\mathcal{N}} \notag \\
    &+ \frac{1}{2} v'''_S v'''_P \big(m_{sp}^2\big)_{\mathcal{N}}, \\
   \mathcal{V}_{\mathcal{N}_{p}} - \mathcal{V}_{\mathcal{N}_s} =& \left[\left(\frac{m_{H^\pm}^2}{4{v'}^2}\right)_{\mathcal{N}_s} - \left(\frac{m_{H^\pm}^2}{4{v''}^2}\right)_{\mathcal{N}_{p}}\right](v_2''v_1' - v_1''v_2')^2 +\frac{1}{4} {v''_P}^2 \big(m_p^2\big)_{\mathcal{N}_s} -\frac{1}{4} {v'_S}^2 \big(m_s^2\big)_{\mathcal{N}_p} , \\
    \mathcal{V}_{\mathcal{N}_{sp}} - \mathcal{V}_{\mathcal{N}_s} =& \left[\left(\frac{m_{H^\pm}^2}{4{v'}^2}\right)_{\mathcal{N}_s} - \left(\frac{m_{H^\pm}^2}{4{v'''}^2}\right)_{\mathcal{N}_{sp}}\right](v_2'''v_1' - v_1'''v_2')^2 +\frac{1}{4} {v'''_P}^2 \big(m_p^2\big)_{\mathcal{N}_s} + \frac{1}{2} v'''_S v'''_P \big(m_{sp}^2\big)_{\mathcal{N}_s}  \notag \\
    &-\frac{1}{4} {v'_S}^2 \big(m_s^2\big)_{\mathcal{N}_{sp}}, \\
    \mathcal{V}_{\mathcal{N}_{sp}} - \mathcal{V}_{\mathcal{N}_p} =& \left[\left(\frac{m_{H^\pm}^2}{4{v''}^2}\right)_{\mathcal{N}_p} - \left(\frac{m_{H^\pm}^2}{4{v'''}^2}\right)_{\mathcal{N}_{sp}}\right](v_2'''v_1'' - v_1'''v_2'')^2 +\frac{1}{4} {v'''_S}^2 \big(m_s^2\big)_{\mathcal{N}_p} + \frac{1}{2} v'''_S v'''_P \big(m_{sp}^2\big)_{\mathcal{N}_p}  \notag \\
    &-\frac{1}{4} {v''_P}^2 \big({m_p}^2\big)_{\mathcal{N}_{sp}} .
\end{align}
From them, it is possible to infer that \textit{if the
$\mathcal{N}, \mathcal{N}_s$ and $\mathcal{N}_p$ minima coexist
in the potential, then the global minimum conserves charge and CP,
but if $\mathcal{N}_{sp}$ also exists, the same may not always be true}.

The minima of the types $\mathcal{S}$ and $\mathcal{P}$ exist only if $m^2_S < 0$ and $m^2_P < 0$. Hence, the analysis of their stability must be done numerically. It is also not possible to analyze the stability of the $\mathcal{SP}$ minima analytically.


\bibliographystyle{JHEP}
\bibliography{ref}

\end{document}